\documentclass[11pt]{article}
\usepackage{graphicx}
\usepackage{amssymb}
\usepackage{amsmath}
\usepackage{graphicx}
\usepackage{amstext}
\usepackage{leftidx}
\usepackage{esint}
\usepackage[utf8]{inputenc}
\usepackage{color}
\usepackage{cite}
\usepackage{hyperref}
\usepackage{multido}

\usepackage{multido}

\makeatletter
\ams@newcommand{\vardot}[2]{%
  {\mathop{#2\kern0pt}\limits^{\vbox to-1.4\ex@{\kern-\tw@\ex@
   \hbox{\normalfont\multido{}{#1}{.}}\vss}}}}
\makeatother
\newcommand{\rL}{\rho_\Lambda}

\newcommand{\CC}{\Lambda}

\newcommand{\rv}{\rho_{\rm vac}}
\newcommand{\Pv}{P_{\rm vac}}
\newcommand{\rvo}{\rho^0_{\rm vac}}

\newcommand{\rco}{\rho^0_{c}}

\newcommand{\nueff}{\nu_{\rm eff}}

\newcommand{\bk}{{\bf k}}
\newcommand{\mpl}{m_{\rm Pl}}
\newcommand{\MPl}{{\cal M}_{\rm Pl}}

\newcommand{\be}{\begin{equation}}
\newcommand{\ee}{\end{equation}}

\newcommand{\ha}{\hat{a}}
\newcommand{\astar}{a_{*}}

\newcommand{\wv}{w_{\rm vac}}

\begin{document}

\hyphenation{theo-re-ti-cal gra-vi-ta-tio-nal theo-re-ti-cally mo-dels}

\begin{center}
\vskip 2mm
{\bf \LARGE  Running vacuum and $H^4$-inflation}

 \vskip 8mm

\textbf{Joan Sol\`a Peracaula$^*$\footnote{Correspondence author: sola@fqa.ub.edu.}, Cristian Moreno-Pulido$^\dagger$ and  Alex Gonz\'alez-Fuentes$^*$}

\vskip 0.5cm
$^*$ Departament de F\'isica Qu\`antica i Astrof\'isica, \\
and   Institute of Cosmos Sciences,\\ Universitat de Barcelona,
Av. Diagonal 647, E-08028 Barcelona, Catalonia, Spain

\vskip0.2cm
$^\dagger$Departament d'Inform\`atica, Matem\`atica Aplicada i Estad\'\i stica, \\
Universitat de Girona, \\
Campus Montilivi
17003  Girona, Spain

\vskip0.5cm


 \vskip2mm

\end{center}
\vskip 15mm

\begin{quotation}
\noindent {\large\it \underline{Abstract}}.
Recent studies of QFT in cosmological spacetime indicate that the speeding up of the present universe may not just be associated with a rigid cosmological term but with a running one that evolves with the expansion rate: $\Lambda=\Lambda(H)$.  This running is inherited from the cosmic evolution of the vacuum energy density (VED),  $\rho_{\rm vac}$,  which is sensitive to quantum effects in curved spacetime that ultimately trigger that running. The VED is a function of the Hubble rate and its time derivatives: $\rho_{\rm vac}=\rho_{\rm vac}(H, \dot{H},\ddot{H},...)$.   Two nearby points of the cosmic evolution during the FLRW epoch are smoothly related as $\delta\rho_{\rm vac}\sim {\cal O}(H^2)$.    In the very early universe, in contrast, the higher powers of the Hubble rate take over  and bring about a period of fast inflation. They originate from quantum effects on the effective action of vacuum, which we compute. Herein we focus on the lowest possible power for inflation to occur: $H^4$.  During the inflationary phase,  $H$ remains approximately constant and very large. Subsequently, the universe enters the usual FLRW radiation epoch.  This new mechanism (`RVM-inflation')  is not based on any supplementary `inflaton' field, it is fueled by pure QFT effects on the dynamical background and is different from Starobinsky's inflation, in which $H$ is never constant.

\end{quotation}

\newpage


\newpage
\newpage
\section{Introduction}
It is an undeniable fact that the standard or concordance model of cosmology, aka $\CC$CDM,   has proven to be a rather successful practical framework for the description of the Universe in the last few decades\,\cite{Peebles1993}. However, it is no less true that a number of serious pitfalls have besieged the $\CC$CDM more recently, which raise serious doubts about its viability as a consistent theory of the cosmological evolution, not even at the mere phenomenological level. Dark matter (DM) is still on the tightrope since it has never been found thus far, a fact that is more than worrisome, as otherwise we cannot understand the dynamical origin of the large-scale structures  that we observe. On the other hand, the core ingredient of the model, namely the cosmological constant (CC), $\CC$,  proves to be an endless source of headaches for theoretical cosmologists.  Despite $\CC$ being introduced into the gravitational field equations by Einstein 108 years ago\,\cite{Einstein1917},  a serious theoretical conundrum is usually associated with that term.  The nature of the problem was put forward by Y. B. Zeldovich\,\cite{Zeldovich1967} half a century later and it goes under the name of   `cosmological constant problem' (CCP)\,\cite{Weinberg89}, being  still in force since then.   In a nutshell, it says that the manyfold successes of quantum field theory (QFT)  in the world of elementary particles appear to be a blatant fiasco in the realm of gravity, the reason being that QFT predicts a value for $\rv$ that is disturbingly much larger than that of the current critical density\,\cite{Weinberg89,Sahni:1999gb,Carroll:2000fy,PeeblesRatra2003,Padmanabhan2003,Copeland2006,Aitchison2009,JSPRev2013,JSPRev2015,JSPRev2022}.

To put it in simple terms, the CCP stems from the fact that the observed  value of the CC, $\Lambda_{\rm obs}$,  should be related to that of the vacuum energy density (VED) in QFT through $\rv=\Lambda_{\rm obs}/(8\pi G)$, and this leads to an extreme fine-tuning between different contributions\footnote{For an informal introduction to the Cosmological Constant Problem, see  \cite{JSPCosmoverse}. For a more formal exposition of the problem along the lines of our presentation, see \cite{JSPRev2022}. }.
More specifically, the problem originates from the mismatch between the huge theoretical contribution triggered by the vacuum fluctuations of the quantum fields  and the tiny observed value  $\rv\sim 10^{-47}\, $GeV.  Theoretically, vacuum fluctuations are expected to be of the order $\sim m^4$ for any quantized field of mass $m$ and are, therefore, far greater than the measured value of $\rv$ -- given the large average mass of the known particles in the standard model of particle physics.  As a consequence, when we put together the sum of all the quartic mass contributions to the VED, this enforces an extremely unreasonable fine-tuning of the parameters. Addressing this fundamental problem is one of the most important tasks of modern theoretical cosmology\,\cite{Weinberg89}. Over the years, the CCP has been dealt with in the literature from many different points of view, although sometimes without taking into account the essential elements that must be involved. A more careful assessment of this issue in the context of QFT in curved spacetime must involve at least two basic ingredients, to wit: the parameter $\rho_\Lambda$ in the Einstein-Hilbert action and the zero-point energy (ZPE) of the quantized fields-- a pure quantum effect. By combining these two elements in the context of quantum field theory in curved spacetime, one obtains the renormalized vacuum energy density.  Furthermore, in the presence of spontaneous symmetry breaking (SSB), one must still account for the contribution from the vacuum-expectation value of the effective potential, which involves in general classical and quantum parts\,\cite{JSPRev2022}. In this paper, however, we will focus only on the ZPE contribution since it is in itself rather cumbersome to deal with in curved spacetime and moreover is of pure quantum nature, which means that is a generic effect in all QFT theories, whether they involve SSB or not. Finally, an important technical ingredient in our approach is that we renormalize the energy-momentum tensor using the method of adiabatic regularization\,\cite{BirrellDavies82,ParkerToms09,Fulling89}, but more specifically we use an off-shell variant of this method which has demostrated to be particularly advantageous to explore the cosmological evolution of the vacuuum energy\cite{JSPRev2022}.
However, these QFT models of the vacuum energy should not be confused with the well-known  ``$\CC(t)$-class'' of time-evolving $\CC$-models,  often referred to as  `Decaying Vacuum Cosmologies', see e.g. \cite{Overduin98} and references therein.  The large  $\CC(t)$-class of phenomenological models has no obvious relation with fundamental physics, but was useful in the past to illustrate the possible effects implied by a time-evolving cosmological term. In a very different vein, the proposal of dynamical scalar field models broke out trying to explain the value of the vacuum energy (believed to be zero) through an automatic adjustment
mechanism in field theory\,\cite{Dolgov82,Abbott85,Banks85, PSW87,Barr87,Ford87,Sola89,Sola90}. Subsequently, the popular class of quintessence models (for a review, see particularly \cite{DEBook} and
\cite{PeeblesRatra2003,Padmanabhan2003,Copeland2006}) emphasized the possible dynamical character of the dark energy (DE) without attempting to compute its value. These models must also be clearly distinguished from the VED option on which we focus here. They are nevertheless useful since many fundamental notions of DE, including the dynamical VED proposal that we endorse in this paper, can mimic quintessence (or phantom DE) behavior.

The renormalization framework of the running vacuum model (RVM) -- see\,\cite{JSPRev2013,JSPRev2015,JSPRev2022} and references therein -- provides a notorious example of framework where the foregoing ideas on the dark energy stemming from the quantum vacuum properties can be implemented in practice within the fundamental framework of  QFT in curved spacetime.  In fact, in recent works, it has been shown that the  CCP can be alleviated in the RVM context, see \cite{CristianJoan2020,CristianJoan2022a,CristianJoan2022b,CristianJoanSamira2023}. The physical outcome is that the measurable value of $\rv$, and hence of  $\CC$,   runs smoothly with the cosmic expansion.  This running, in fact, can be thought of as a renormalization group (RG) running. To construct the physically renormalized VED  we have to start from the higher order bare action in QFT in curved spacetime\,\cite{BirrellDavies82,ParkerToms09} and produce an appropriately renormalized one. At that point the VED can be constructed from the mentioned renormalized term $\rL (M)$ in the Einstein-Hilbert part of the action and the renormalized ZPE, which also depends on $M$,  where $M$ is the renormalization scale.  We may write the combined quantity very qualitatively as follows:  ${\rm VED}=\rho_\Lambda+{\rm ZPE}$.
The  renormalized result, however,  still requires a  physical interpretation since it depends on the renormalization scale $M$.  Its  appearance  is characteristic of the renormalization procedure in QFT owing to
the intrinsic breaking of scale invariance by the quantum effects.  Despite the fact that the  full effective action is independent of $M$, different parts of it are indeed $M$-dependent. In particular, the vacuum effective action induced by quantum effects of the quantized matter fields, $W_{\rm eff}$\,\cite{BirrellDavies82,ParkerToms09},  is explicitly dependent on the renormalization scale $M$.   For a physical interpretation of the theory, an adequate choice of $M$ at the end of the renormalization program  is mandatory.  Following the standard practice in ordinary gauge theories,  the choice of  the renormalization scale  is  made near the typical energy of the process.  In the RVM framework  it is suggested  to choose $M$   around the characteristic energy scale of cosmic spacetime at any given moment, namely the value of the Hubble rate $H$ in natural units at each cosmic epoch, see \cite{CristianJoan2020,CristianJoan2022a,CristianJoan2022b,CristianJoanSamira2023} for full details and \cite{JSPRev2022} for a review.

The quantum vacuum in the RVM context is therefore dynamical rather than just stuck at a rigid value. It means that there is no such thing as a rigid cosmological constant within a QFT explanation of the expanding universe. No less remarkable is that the theory  appears to be free from fine tuning issues caused by the aforesaid quartic mass contributions $\sim m^4$ of the quantum fields. These fine tuning troubles have been ascribed too often and exclusively to the cosmological constant $\CC$, but they are actually not present in the RVM framework\,\cite{CristianJoan2020,CristianJoan2022a}, a fact which constitutes an important achievement of the RVM.

The headaches with the $\CC$CDM involve not only high brow fundamental theoretical conundrums such as the CCP, they also  impinge on very practical and ground level issues of modern cosmology, such as  the current cosmological tensions between the standard model predictions and different sorts of data. In particular,  tensions with the local measurement of the current Hubble parameter $H_0$ as well as with the growth of  large scale structures.  The $H_0$-tension involves a  serious disagreement ($\sim 5 \sigma$ c.l.)  between the value of $H_0$ inferred from  CMB observations, which make use of fiducial $\CC$CDM cosmology, and the corresponding value extracted from the distance ladder measurements.  The growth tension, on the other hand,  is related with the overproduction  of large scale structure in the late universe as predicted by the $\CC$CDM in comparison with actual measurements. The tension here is moderate ($\sim 2-3\sigma$)  but persistent.  See e.g. \cite{Abdalla:2022yfr,Perivolaropoulos:2021jda} and references therein for a comprehensive review of these tensions.  
 Interestingly enough, it turns out that the RVM framework may also provide a clue for the resolution of the aforementioned practical problems. The predicted running law $\delta\rv\sim H^2$ of the VED in the current universe has been successfully tested against the global cosmological observations in different studies, e.g. in the works\,\cite{SolaPeracaula:2021gxi} and the older ones\,\cite{Sola:2015wwa,SolaPeracaula:2016qlq}. In the former, tensions are substantially alleviated. Notable are also the RVM-inspired proposals\,\cite{CompositeDE,BDRVM,BDRVM2}, which also provide tantalizing solutions to these tensions from different perspectives\,\cite{SolaPeracaula:2024iil}.
 See also\,\cite{Akarsu:2021fol,Akarsu:2023mfb,Gomez-Valent:2023uof,Giare:2025pzu,Akarsu:2025gwi,Soriano:2025gxd,Dwivedi:2024okk,Anchordoqui:2023woo} for alternative proposals and discussions on the tensions and nature of DE.

A variety of dynamical DE models of various sorts have recently appeared being spurred by the urgent need to fix the cosmological tensions: see, for instance,  models of Early DE\, \cite{Gomez-Valent:2021cbe,Kamionkowski:2022pkx,Poulin:2023lkg}, Emergent DE\,\cite{Yang:2021eud}, interacting DE\,\cite{Gomez-Valent:2022bku,Wang:2024vmw} as well as modified $f(R)$ \cite{Sotiriou:2008rp,Capozziello:2011et} and teleparallel $f(T)$ gravities\,\cite{Cai:2015emx}. Scalar-tensor theories\,\cite{Horndeski:1974wa,Kobayashi:2019hrl,Clifton:2011jh}, on the other hand, can also be used to describe dynamical DE. Notably, Brans-Dicke theories\,\cite{BransDicke1961}
with a cosmological term can also mimic dynamical DE and can help curing the tensions\,\cite{BDRVM,BDRVM2}.


Apart from the impact on the physics of the current universe, the RVM also has predictive power for the dynamics of the very early universe, in particular for the  inflationary stage.  Although the VED evolves as $\sim H^2$ in the post-inflationary epoch, as noted above,  the RVM predicts that inflation is driven by higher powers of the expansion rate beyond $H^2$.  Spurred by the phenomenological success of the RVM in describing the late universe,  the present study also explores the implications of the RVM for the physics of the very early universe, where higher (even) powers of the Hubble rate $\sim {\cal O}(H^N)$ ($N=4,6,...$) may bring about inflation. In contrast to previous phenomenological approaches, see e.g. \cite{BLS2013,Perico:2013mna,Sola:2015csa,Yu2020}, where no formal justification is given for these higher powers in the structure of the VED, here we account for their existence  in the context of  QFT in curved spacetime and focus on the inflationary effect prompted by $H^4$.    We note that such higher  powers  can also be attained  from the VED structure of dynamically broken supergravity models with gravitino condensation\,\cite{Basilakos:2015yoa}, which emerge  from the low energy
limit of string theories, see \,\cite{ReviewNickJoan2021,PhantomVacuum2021} and  \cite{BasMavSol,NickPhiloTrans,Gomez-Valent:2023hov,Dorlis2024} and the forthcoming comprehensive review\,\cite{NickJoan_PR}.  The type of inflation produced by the $H^4$-term --- and, in general, by  higher order  (even) powers of $H$ --- is characteristic of  `RVM-inflation'.  We should remark that the latter  follows a different pattern as compared to Starobinsky's inflation\,\cite{Starobinsky:1980te}, although graceful exit is still perfectly granted -- see\,\cite{JSPRev2015,Basilakos:2015yoa} for a comparison with Starobinsky's inflation.  Being the RVM contribution nonvanishing for $H=$const.  and taking into account that the Starobinsky-like higher order terms just vanish in such a regime, it is reasonable to expect that for large values of the Hubble rate the  $\sim H^4$  power (triggered by quantum effects) prove to be the main character  at the inflationary scale. In this sense, the RVM  approach is genuinely new. The unified description of the cosmic evolution that is  gathered  in this context could help changing our picture of inflation  into a more RVM-like one.

All in all, the presence of the higher powers of the Hubble rate  in the early universe can be very important from different perspectives. One more example, as noted in\,\cite{Mavromatos:2020crd}, is that they could help evading the possible trouble of string theories with the  `swampland' criteria on the  impossibility to construct metastable de Sitter vacua, which if so it would forbid the existence of de Sitter solutions in a low energy effective theory of quantum gravity.  Besides, the existence of the $H^4$- terms does not depend on picking out a particular potential for the scalar field, as  no potential  will be  introduced  in the RVM context. Thus, the RVM approach, whether in QFT or stringy formulations, may provide a self-consistent framework for inflation within fundamental physics and with minimal assumptions.   In the present study, however, we will focus exclusively on the QFT version of RVM-inflation.

The various materials of interest presented in this paper can be summarized as follows.  In Sec.\ref{sec:VEDflat} we discuss the relation between the concepts of vacuum energy and cosmological constant. In Sec. \ref{ZPEScalar}, we compute the renormalized energy-momentum tensor of vacuum using the powerful method of the effective action. From this result, we derive a cosmological evolving VED that is remarkably free from quartic mass contributions $\sim m^4$, a fact that helps alleviate the fine-tuning problem as part of the CCP. Then we exploit the consequences of such a smooth running of the VED in both the late and early universe. In the former, we derive a mildly evolving $\CC(H)$-term which deviates by $\sim H^2$ from the rigid $\CC$ value of the concordance model. We analyze also the effective equation of state of the running vacuum, which turns out to mimic quintessence or phantom DE. Finally, we present a devoted study of the implications of the RVM in the early universe  (Sec. \ref{sec:RVMInflation}), and show that inflation can be triggered by a $\sim H^4$ power. This leads to an alternative inflationary mechanism, called `RVM-inflation'. We discuss in detail its thermodynamical properties and show that it can cure basic problems of the $\CC$CDM framework concerning, for instance, the large entropy of the current universe.  The final discussion and conclusions of our paper are given in Sec.\ref{sec:conclusions}.

\section{Vacuum Energy and Cosmological Constant}\label{sec:VEDflat}

Vacuum energy and cosmological constant are different physical concepts. It is important to emphasize this fact since in the extensive literature on this subject over the years one finds that these two concepts are sometimes identified and this may cause some confusion. A more careful consideration shows that the physical relationship between these concepts is possible only in curved spacetime. Let us first consider Minkowski spacetime. In this context the calculation of the VED  for the case of a single free scalar field $\phi$ is completely standard, usually performed using  Minimal Subtraction (MS) and dimensional regularization (DR). The result is known since long ago and it is even discussed in old textbooks of QFT\,\cite{Brown1992} using path integral methods, or  alternatively from a direct calculational approach to the ZPE\,\cite{Akhmedov:2002ts,Ossola:2003ku}.  At one-loop order it leads to the renormalized result (see also\,\cite{JSPRev2013,JSPRev2022} for an expanded discussion):
\begin{equation}\label{VEDMink}
\rv=\rL(\mu)+\frac{m^4}{64\pi^2}\,\left(\ln\frac{m^2}{\mu^2}+C_{\rm
vac} \right)\,.
\end{equation}
In this equation,  $\rL(\mu)$ is the renormalized additive constant term in the action\footnote{Despite the notation, the term $\rL$ has no a priori relation  with the bare cosmological term of a gravitational action since there is no gravity in this (flat spacetime) context. See, however, the next section. } and $\mu$ is the conventional  't Hooft's mass unit of DR\,\cite{Collins84}.
The two-loop calculation (involving self-interaction of the scalar field) is also available and leaves formally unmodified the above one-loop result\,\cite{Brown1992},  but this nicety does not change the situation that we wish  to stand out here.  The result for an elementary fermion field is also well-known (e.g. if it is uncolored it just amounts to multiply  the coefficient of the second term on the \textit{rhs} of Eq.\,\eqref{VEDMink} by a factor of $-4$), but for definiteness we will focus here on scalar fields.
The mentioned second term is the renormalized ZPE at one-loop in the MS.  An arbitrary constant  $C_{\rm vac}$  is left  undetermined after canceling the pole  in $n=4$ spacetime dimensions  since the pole cancellation  in the MS can be performed by including any additive constant.  As it is obvious, Eq.\,\eqref{VEDMink} has no connection with the expanding universe, in particular it does not depend on its expansion rate $H(t)$ or any other cosmological parameter.  This point is crucial and reminds us of another important aspect in dealing with the CCP: we should  not thoughtlessly associate VED with a cosmological constant (CC), as done too often in the literature.  The former may exist in Minkowski spacetime, as given e.g. in Eq.\,\eqref{VEDMink}, whereas the latter can exist only in the context of Einstein's equations in curved spacetime and hence in the presence of gravity.    However,  the above result  illustrates in a very manifest way that a calculation of the VED in flat spacetime, even after renormalization,  has no impact whatsoever on the physics of the CC since the latter cannot be defined in Minkowskian spacetime!  In fact, $\CC\neq 0$ is inconsistent with the solution of Einstein's equations in that spacetime. Thus,  in order to make contact with the physical $\CC$, we need to move to curved spacetime and compute the renormalized VED in an appropriate renormalization framework, namely one imbued with more physical meaning than just the formal renormalization performed within the MS scheme, in which the scale $\mu$ is a pure artifact having no obvious relation with the physics of the cosmological spacetime.  By just cursory setting $\mu$ equal to $H$ in the above flat spacetime result (as done sometimes in the literature so as to inject some cosmological physics  in the last minute) makes actually no sense at all since - as  we should reiterate  --  the VED in Minkowski spacetime has no relation whatsoever with the CC, see \cite{CristianJoan2022a,JSPRev2022} for a detailed discussion.  The setting $\mu=H$ can only make sense in the context of calculations performed \textit{ab initio}  in cosmological spacetime, which is when the two concepts under discussion get logically entangled. This is what we do right next.

\section{Energy-Momentum Tensor and Effective Action for a Non-Minimally Coupled Scalar Field in QFT}\label{ZPEScalar}
Consider a quantized scalar matter  field $\phi$ of mass $m$ non-minimally coupled to curvature in Friedmann-Lema\^\i tre-Robertson-Walker (FLRW) spacetime.  The corresponding Einstein-Hilbert (EH) action reads\footnote{Our metric and curvature conventions are as in \cite{CristianJoan2022a}, see particularly Appendix A of that reference.}
\begin{equation}\label{eq:EH}
S_{\rm EH+\phi}=  S_{\rm EH}+S_{\phi}=\int d^4 x \sqrt{-g}\,\left(\frac{1}{16\pi G}\, R  - \rL\right) + S_{\phi}\,,
\end{equation}
and the scalar field action is
\begin{equation}\label{PhiAction}
S_\phi=-\int d^4x \sqrt{-g}\left(\frac{1}{2}g^{\mu\nu}\partial_\mu \phi \partial_\nu \phi+\frac{1}{2}\left(m^2+\xi R\right)\phi^2\right)\,.
\end{equation}
We pointed out in the previous section that the cosmological term is  physically meaningful only in curved spacetime.  Its observed value $\CC_{\rm obs}$ becomes naturally linked to the VED $\rv$ through Einstein's equations: $\rv=\Lambda_{\rm obs}/(8\pi G)$.
However, even in this gravitationl context we should not confuse  the bare parameter  $\rL$ in the EH action with the physical $\rv$, where the former is related in a similar way with the bare values of $\CC$  and $G$ in a gravitational context. The corresponding connection with physical quantities is not immediate at this point, and it will only emerge upon properly renormalizing the theory.   Einstein's equations follow from the standard variation of the total action $S_{\rm EH+\phi}$  with respect to the metric:
\begin{equation} \label{EinsteinEqs}
\frac{1}{8\pi G}G_{\mu \nu}=-\rho_\Lambda g_{\mu \nu}+T_{\mu \nu}^{\phi}\,,
\end{equation}
where  $G_{\mu\nu}=R_{\mu\nu}-(1/2) g_{\mu\nu} R$  is the usual Einstein tensor.  The above are classical field equations, which do not yet incorporate quantum effects. This will be done upon quantization.

\subsection{From Classical to Quantum Field Theory }\label{sec:Classical}

The classical energy-momentum tensor (EMT) for $\phi$ is obtained from the functional variation of the matter action \eqref{PhiAction} as follows:
\begin{equation}
\begin{split}\label{EMTScalarField}
T_{\mu\nu}^\phi &= -\frac{2}{\sqrt{-g}}\frac{\delta S_\phi}{\delta g^{\mu\nu}}=\left(1-2\xi\right) \partial_\mu \phi \partial_\nu \phi +\left(2\xi-\frac{1}{2}\right)g_{\mu \nu}\partial^\sigma \phi \partial_\sigma \phi\\
&-2\xi\phi \nabla_\mu \nabla_\nu\phi + 2\xi g_{\mu\nu}\phi \Box \phi+\xi G_{\mu\nu}\phi^2-\frac{1}{2}m^2 g_{\mu\nu}\phi^2\,.
\end{split}
\end{equation}
It is well known that the action $S_\phi$ given by \eqref{PhiAction} becomes (locally) conformal invariant in the massless limit and for the value $\xi=1/6$ of the non-minimal coupling.  In our case, however,  the value of $\xi$ will not be fixed a priori and we will not assume massless particles since some of the main effects we aim at hinge on having massive matter fields\cite{CristianJoan2020,CristianJoan2022a}.  Varying now the action with respect to the scalar field leads to the  Klein-Gordon (KG) equation with non-minimal coupling:
\begin{equation}\label{KG}
\left(\Box-m^2-\xi R\right)\phi^2=0\,,
\end{equation}
where $\Box\phi=g^{\mu\nu}\nabla_\mu\nabla_\nu\phi=(-g)^{-1/2}\partial_\mu\left(\sqrt{-g}\, g^{\mu\nu}\partial_\nu\phi\right)$.
As indicated, we perform the calculation in cosmological (FLRW) spacetime. We assume a flat three-dimensional metric and for convenience, we use the conformal frame  $ds^2=a^2(\tau)\eta_{\mu\nu}dx^\mu dx^\nu$, with  $\eta_{\mu\nu}={\rm diag} (-1, +1, +1, +1)$  the Minkowski  metric in our conventions,  $a(\tau)$ is the scale factor and  $\tau$  the conformal time. Differentiation with respect to $\tau$ will be denoted by a prime, so, for example, $\mathcal{H}\equiv a^\prime/a$ is the corresponding Hubble function in conformal time. Although calculations will be performed using the conformal metric, our final results will be rendered in terms of the usual Hubble function $H(t)=\dot{a}/a$ in cosmic time $t$ (where a dot denotes differentiation with respect to $t$).  Since $d\tau=dt/a$, we have $\mathcal{H}=a H$, $\mathcal{H}^\prime=a^2(H^2+\dot{H})$,
$\mathcal{H}^{\prime\prime}=a^3\left(2H^3+4 H\dot{H}+\ddot{H}\right)$, etc., which are useful relations for the mentioned conversion.

The above equations hold good in classical field theory. However, in QFT, $\phi$ is a quantized matter field contributing also with quantum fluctuations $\delta\phi$.  In this case,  it is convenient to separate the fluctuations from the background part:
 \begin{equation}
 \phi\left(\vec{x},\tau \right) = \phi_b (\tau)+\delta\phi \left(\vec{x},\tau \right)\,.
 \end{equation}
Notice that the background field $\phi_b (\tau)$ is assumed to be spatially homogeneous.
The fluctuating part is not spatially homogeneous and can be decomposed in Fourier frequency modes:
\begin{equation}\label{FourierDecomposition}
\delta \phi(\tau,{\bf x})=\frac{1}{(2\pi)^{3/2}a}\int d^3{k} \left[ A_\bk e^{i{\bf k\cdot x}} h_k(\tau)+A_\bk^\dagger e^{-i{\bf k\cdot x}} h_k^*(\tau) \right]\,.
\end{equation}
Upon quantization, the Fourier expansions are promoted
to operators in the Heisenberg representation, on which we impose the canonical commutation
relations. These are encoded in those satisfied by the creation and annihilation operators, $A_k$ and $A_k^\dagger$:
\begin{equation}
[A_\bk, A_\bk'^\dagger]=\delta({\bf k}-{\bf k'}), \qquad [A_\bk,A_ \bk']=0\,. \label{CommutationRelation}
\end{equation}
Using these relations, the KG-equation \eqref{KG} in terms of frequency modes can be put as
\begin{equation} \label{KGmodes}
h^{\prime\prime}_{k}+\Omega_k^2(\tau)h_k=0\,,
\end{equation}
where $\Omega_k^2\equiv k^2+a^2 m^2+a^2\left(\xi-1/6\right)R$, with $k\equiv|{\bf k}|$ the modulus of the
comoving momentum (the physical momentum being $k/a$). Since the effective frequency $\Omega_k(\tau)$ is not constant, the above linear differential equation corresponds to an anharmonic oscillator and does not possess a close analytic solution, except for very particular cases, such as e.g. the massless limit with minimal coupling ($\xi=0$) -- a situation which is far from our main interests.  So in general, one must resort to what is called an adiabatic series expansion, which is essentially a WKB-type solution\,\cite{BirrellDavies82,ParkerToms09,Fulling89}. To this aim, one introduces a phase-integral ansatz for the mode function:
\begin{equation}\label{eq:phaseintegral}
h_k (\tau)=\frac{1}{\sqrt{2W_k (\tau)}}\exp \left(-i\int^\tau W_k (\tilde{\tau})d\tilde{\tau}\right)\,,
\end{equation}
which is normalized through the Wronskian condition
 $h_k^{} h_k^{*\prime}-h_k^\prime h_k^* = i$. The above template is motivated by the fact that for constant $\Omega_k$ (i.e. independent of time), it provides the exact solution for positive frequency modes.
Proceeding in this way, we have traded the original mode function $h_k$ for the new function $W_k$, which satisfies a nonlinear (WKB-type) differential equation:
\begin{equation}\label{Non-LinDiffEq}
W_k^2=\Omega_k^2 -\frac{1}{2}\frac{W_k^{\prime \prime}}{W_k}+\frac{3}{4}\left( \frac{W_k^\prime}{W_k}\right)^2\,.
\end{equation}
For a slowly varying effective frequency $\Omega_k (\tau)$ one can solve this equation perturbatively with the help of an asymptotic series which can be  organized through adiabatic orders. It is obvious that the suitability of the adiabatic expansion is linked to the smallness of the ratio $|\Omega'_k (\tau)/\Omega^2_k (\tau)|\ll1$. This approach constitutes the basis for adiabatic regularization\,\cite{BirrellDavies82,ParkerToms09,Fulling89}, which is essentially an expansion in the number of time derivatives; if the time evolution is slow (measured by the aforementioned ratio), the series converges faster. Each time derivative increases by one unit the adiabatic order and
the ansatz solution for $W_k$ can be written as
\begin{equation}\label{eq:WKBseries}
W_k=\omega_k^{(0)}+\omega_k^{(2)}+\omega_k^{(4)}+\omega_k^{(6)}+\cdots,
\end{equation}
in which the superscript indicates the adiabatic order. If the expansion is carried out up to $Nth$ adiabatic order, the vacuum state annihilated by all ladder operators $A_\bk$ satisfying \eqref{CommutationRelation} is known as the $Nth$-order adiabatic vacuum. The vacuum expectation values (VEV's) computed in the adiabatic approach always refer to that vacuum state at the given order. It is, of course, an approximate vacuum state. One can
see immediately that the adiabatic expansion in the cosmological context ends up as an expansion in powers of H and its time
derivatives\,\cite{CristianJoan2020,CristianJoan2022a,CristianJoan2022b,CristianJoanSamira2023}. Not surprisingly, only even adiabatic orders ($N=0,2,4,...$) are allowed owing to the requirement of general covariance of the solution.  This is confirmed by explicit calculation, where all odd adiabatic orders are absent (see the aforementioned papers). Finally, we should point out that the WKB expansion \eqref{eq:WKBseries} is an asymptotic series and therefore it is understood that it should be truncated to low adiabatic orders, beyond which its convergence gets progressively degraded.

As is well-known, QFT in curved spacetime is a quantum theory of fields, not particles. The presence of the nontrivial modes satisfying Eq.\eqref{KGmodes}, or equivalently the nonlinear WKB one \eqref{Non-LinDiffEq}, already indicates that particles with definite frequencies cannot be strictly defined in a curved
background. So the physics actually resides in the fields and in particular in the properties of the EMT. Now since we are specifically addressing the effect of the quantum fluctuations of the quantized matter fields, we need to focus on computing just the VEV of the EMT, which we may call the `vacuum EMT' for short. This quantity depends on bilinears of the fluctuation $\delta\phi$  and its time derivatives\,\cite{CristianJoan2020}. Being a composite operator made of operator products, we expect the VEV to be divergent in QFT. Hence, to extract physical results on the vacuum effects, renormalization is mandatory.  The Fourier expansion given in \eqref{FourierDecomposition} is to be inserted  in \eqref{EMTScalarField} using the commutation relations \eqref{CommutationRelation}.
The resulting vacuum EMT involves integration over all modes,  $\int\frac{d^3k}{(2\pi)^3}(...) $, which yields UV-divergent integrals up to fourth adiabatic order (in $n=4$ spacetime dimensions). Adiabatic orders higher than $4$ decay sufficiently fast at large momentum $k$ (short distances) to make the corresponding integrals convergent\,\cite{BirrellDavies82}. However, unavoidably, an UV-divergent part of the vacuum EMT must be dealt with, and this means that we need to renormalize it by appropriately subtracting the first four (UV-divergent) adiabatic orders.  In \cite{CristianJoan2020} the following ``off-shell subtraction prescription''   was proposed to renormalize the vacuum EMT:
\begin{equation}\label{RenormalizedEMTScalar}
\left\langle T_{\mu\nu}^{\delta\phi} \right\rangle_{\rm ren} (M)=\left\langle T_{\mu\nu}^{\delta\phi} \right\rangle (m)-\left\langle T_{\mu\nu}^{\delta\phi}\right\rangle^{(0-4)}(M).
\end{equation}
This specific form will be referred to  as off-shell ARP (adiabatic regularization prescription). Being off-shell, it is an extended version of the original (on-shell) procedure\,\cite{BirrellDavies82,ParkerToms09,Fulling89}. In practice, ARP implements both regularization and renormalization of the EMT at the scale $M$.
The superscript $(0-4)$  in the second term on the \textit{rhs} of \eqref{RenormalizedEMTScalar} refers to the (UV-divergent) orders being subtracted, while the first term is the on-shell value. The latter can be computed, in principle, to any desired adiabatic order. By keeping the subtraction scale $M$ generic, however, we can test the evolution of the VED with $M$.  In fact, having a floating scale $M$ in QFT is characteristic of the renormalization group analysis. In the previous section, the alternative scale $\mu$ was used for Minkowski space calculations in the MS scheme with DR, but as pointed out there, it is not convenient in cosmology.  In the present curved spacetime context, in contrast, the above subtracting procedure is more physical and $M$ can eventually be identified with the Hubble rate at the end of the calculation\,\cite{CristianJoan2020,CristianJoan2022a}
\footnote{Off-shell renormalization is actually the clue to our approach\cite{CristianJoan2020,CristianJoan2022a}, as is also the case in other QFT contexts.  For example, the entire QCD theory of strong interactions is renormalized off-shell since the quarks do not participate on-shell in their interactions with gluons. Also, in quantum electrodynamics it allows to discuss the renormalization group running of the fine structure constant, whose confirmation was a major triumph of RG theory. As a matter of fact, off-shell renormalization is completely natural in cosmology if we take into account that the characteristic energy parameter $H$ (n natural units) during most of the cosmological evolution is certainly much smaller than the average mass of any known particle. The exception is during the inflationary period, which we deal in detail in Sec.\ref{sec:RVMInflation}.}. However, rather than renormalizing the vacuum EMT directly by means of the above recipe, as it was done in\,\cite{CristianJoan2020}, we may alternatively compute the renormalized vacuum effective action $W$ within the off-shell ARP framework, which also depends on the scale $M$\,\cite{CristianJoan2022a}, and from it we can  extract the renormalized vacuum EMT. We do this in the next section.

\subsection{Vacuum Effective Action and Its Adiabatic Renormalization }\label{sec:vacuumEA}
The effective action of vacuum, $W$, accounts for the quantum effects from the quantized matter fields. It can be obtained from the DeWitt-Schwinger expansion by integrating out the vacuum fluctuations of these fields\,\cite{BirrellDavies82,ParkerToms09}.  From the knowledge of $W$, one can then derive the VEV of the  EMT,  $\left\langle T_{\mu\nu} \right\rangle$ (i.e. the `vacuum EMT') by computing the usual metric functional derivative, but now using the vacuum effective action:
\begin{equation}\label{eq:DefWeff}
\langle  T_{\mu\nu}^{\delta\phi}\rangle=-\frac{2}{\sqrt{-g}} \,\frac{\delta W}{\delta g^{\mu\nu}}\,.
\end{equation}
This method constitutes an alternative path to the mode expansion for the renormalization of the EMT through the prescription \eqref{RenormalizedEMTScalar}, see\,\cite{CristianJoan2020,CristianJoan2022a} for more details. Let us summarize the procedure. At one loop order (the only one available for the free theory), the value of $W$ is obtained from the trace of the logarithm of the inverse Green's function in curved spacetime.  More specifically,
\begin{equation}\label{eq:EAW}
\begin{split}
W= &\frac{i\hbar}{2}Tr \ln (-G_F^{-1})=\frac{i\hbar}{2}Tr \ln (-G_F)^{-1}= -\frac{i\hbar}{2}Tr \ln (-G_F)\\
=&  -\frac{i\hbar}{2} \int d^4 x \sqrt{-g}\lim\limits_{x\to x'} \ln\left[ -G_F(x,x')\right]\equiv \int d^4 x \sqrt{-g}\, L_W\,.
\end{split}
\end{equation}
The last equality defines the  Lagrangian density $\sqrt{-g}\, L_W$ associated with the quantum vacuum effective action, i.e. that part of the full Lagrangian which encodes the quantum effects from the vacuum fluctuations of the quantized matter fields.   We have kept $\hbar$ explicitly here  only to emphasize the pure quantum character of the above action, but we set $\hbar=1$  henceforth.

In order for the Lagrangian of the theory to be renormalizable, we must include not only the EH terms indicated in Eq.\eqref{eq:EH} but also the usual higher derivative (HD) geometric terms. Hence the full classical part of the Lagrangian takes on the form
\begin{equation}\label{eq:LEHHD}
\begin{split}
L_G^{\rm cl.}=L_{EH}+L_{HD}
=&-\rho_\Lambda+\frac{1}{2}\MPl^2  R+\alpha_1 C^2+\alpha_2 R^2+\alpha_3 E+\alpha_4\Box R\,.
\end{split}
\end{equation}
Here, $\MPl^2\equiv 1/\left(8\pi G\right)=\mpl^2/(8\pi)$ is the (reduced) Planck mass squared, with $\mpl$ the usual Planck mass.
The HD terms, which carry dimensionless coefficients $\alpha_i$, are conveniently grouped using the square of the Weyl tensor ($C^2$) and the Euler density ($E$) plus a total derivative\,\cite{BirrellDavies82, CristianJoan2022a}. The classical Lagrangian \eqref{eq:LEHHD} is not yet the full Lagrangian relevant to our considerations, of course, but is indispensable for the renormalization process. We must still add up to it the effective quantum vacuum action, $L_W$, which at this point is divergent and hence requires renormalization. But before doing that, we need to find the bare $L_W$ Lagrangian from the effective action.

To compute $W$  from \eqref{eq:EAW}, we must solve the Green's function equation in curved spacetime\begin{equation}\label{KGPropagatorOffShell}
\left(\Box_x-M^2-\Delta^2-\xi R(x)\right)G_F(x,x^\prime)=-\left(-g(x)\right)^{-1/2}\delta^{(n)}(x-x^\prime)\,,
\end{equation}
where $\delta^{(n)}$ is the Dirac $\delta$-distribution in $n$ spacetime dimensions. For all practical purposes in our work, $n=4$. It is nevertheless convenient to keep $n$ general since it allows to use DR  for regularizing the UV divergences. Such an auxiliary use of DR here is, however,  completely independent of the renormalization procedure that we subsequently follow, see below.  Obviously an exact solution of \eqref{KGPropagatorOffShell} is not generally possible, but $G_F$ can be found by means of an adiabatic expansion, in complete analogy with the alternate method mentioned in the previous section, which is based on the mode expansion.  Notice that the key for the adiabatic solution of the above equation is the presence of the floating scale $M$ as well as of the important quantity
\begin{equation}\label{eq:Delta2}
\Delta^2\equiv m^2-M^2\,.
\end{equation}
This quantity must be dealt with as being of  adiabatic order $2$ \cite{CristianJoan2022a}.  In fact, considering that the term $\xi R$ in \eqref{KGPropagatorOffShell} is also of adiabatic order $2$, the combination $\Delta^2+\xi R$ is then an adiabatic block of the same order.  Now since the adiabatic order of the terms must be hierarchically respected throughout the expansion, and bearing in mind that the mass scale $M$ is of adiabatic order zero while the special quantity $\Delta^2$ is of adiabatic order $2$, the solution to Green's function equation \eqref{KGPropagatorOffShell} from a consistent adiabatic expansion is different from the solution to the corresponding on-shell equation with $\Delta^2=0$.

For the explicit calculation, one starts from the standard DeWitt-Schwinger expansion of the effective action, which is obtained by computing the curved spacetime propagator $G_F$. In practice, this means to adiabatically expand the solution of Eq.\,\eqref{KGPropagatorOffShell} since an exact approach is not feasible. The calculation is rather cumbersome and the details for the on-shell case can be found in \cite{BirrellDavies82,ParkerToms09}. Once the propagator $G_F$ is found, one must apply Eq.\,\eqref{eq:EAW} in order to identify the effective Lagrangian $L_W$. In the off-shell case one proceeds similarly but now the result involves an explicit dependence on the renormalization scale $M$ and extra terms which depend on the quantity $\Delta^2$ \cite{CristianJoan2022a,FerreiroNavarroSalas2019}. The final result for the effective Lagrangian of the vacuum fluctuations is the following:

\begin{equation}\label{eq:effLagrangian}
L_W=\frac{\mu^{4-n}}{2(4\pi)^{n/2}} \sum_{j=0}^\infty \hat{a}_j (x) \int_0^\infty (is)^{j-1-n/2}e^{-iM^2 s}ids
=\frac{1}{2(4\pi)^{2+\frac{\varepsilon}{2}}}\left(\frac{M}{\mu}\right)^{\varepsilon}\sum_{j=0}^\infty \hat{a}_j (x) M^{4-2j}\Gamma \left(j-2-\frac{\varepsilon}{2}\right)\,.
\end{equation}

Recall that we use DR to regularize the divergences and we defined  $\varepsilon\equiv n-4$. The limit $\varepsilon \rightarrow 0$ is understood. As usual,  $\Gamma$ is Euler's gamma function and $\mu$ is  the aforementioned 't Hooft's mass unit to keep the effective Lagrangian with natural dimension  $+4$ of energy  in $n$ spacetime dimensions, the final results being independent of $\mu$, which is an unphysical parameter\footnote{In fact, our final renormalized result depends on $M$ only, not on the auxiliary $\mu$ introduced for DR regularization purposes. In contrast, in the approach of\,\cite{KohriMatsui2017}, which lacks of our subtraction prescription at $M$, the final results  still carry explicit $\mu$-dependence and calculations lead to the unwanted $\sim m^4$ contributions responsible for extreme fine-tuning in the CCP.}.   The sum over   $j=0,1,2,...$  involves  the even adiabatic orders only.  The presence of $\Delta^2$ modifies the  DeWitt-Schwinger coefficients $\hat{a}_i (x)$. Up to fourth adiabatic order the new coefficients read

\begin{equation}\label{eq:ModifDWScoeff}
\begin{split}
&\hat{a}_0 (x)=1=a_0 (x),\\
&\hat{a}_1 (x)=a_1(x)-\Delta^2=-\left(\xi-\frac{1}{6}\right)R-\Delta^2 ,\\
&\hat{a}_2 (x)=a_2(x)+\frac{\Delta^4}{2}+\Delta^2 R \left(\xi-\frac{1}{6}\right)=\frac{1}{2}\left(\xi-\frac{1}{6}\right)^2R^2+\frac{\Delta^4}{2}+\Delta^2 R \left(\xi-\frac{1}{6}\right)-\frac{1}{3}Q^\lambda_{\ \lambda}\,,
\end{split}
\end{equation}
with
\begin{equation}\label{eq:traceQ1}
\frac{1}{3}{Q^\lambda}_\lambda \equiv-\frac{1}{120}C^2+\frac{1}{360}E+\frac{1}{6}\left(\xi-\frac{1}{5}\right)\Box R\,.
\end{equation}
The coefficients  $a_i(x)$ are the ordinary  DeWitt-Schwinger coefficients for  $\Delta=0$ (on-shell expansion).   The effective Lagrangian \eqref{eq:effLagrangian} is manifestly UV-divergent  since Euler's $\Gamma$-function is divergent for $j=0,1,2$  in  $n=4$ spacetime dimensions. Therefore, renormalization is required.  We avoid using the MS scheme in this context; instead, we utilize the off-shell ARP  adapted to the effective action.  In fact,  we define the renormalized vacuum effective Lagrangian at the
renormalization point M as follows:
\begin{equation}\label{eq:LWrenormalized}
L_W^{\rm ren}(M )= L_W (m)-L_W^{(0-4)}(M)\equiv  L_W (m)-L_{\rm div} (M)\,,
\end{equation}
where  $L_{\rm div}(M)\equiv  L_W^{(0-4)}(M)$ is the divergent part.  Notice that  $L_W (m)$  and $L_{\rm div} (M)$ are both divergent, but the former may involve the full DeWitt-Schwinger expansion at any desired order, whereas the latter is computed only up to order $4$.  This subtraction prescription, which is  performed at the level of the effective Lagrangian, is the exact analogue of the off-shell ARP definition \eqref{RenormalizedEMTScalar} for the EMT and it suffices to make $L_W^{\rm ren}(M)$ finite.  The pole terms (which appear in the limit $\varepsilon\to0$) exactly cancel out in Eq.\,\eqref{eq:LWrenormalized}, leaving the following finite result:
\begin{equation}\label{eq:LWrenM}
\begin{split}
L_W^{\rm ren}(M)=\delta \rho_\Lambda(M)-\frac{1}{2}\delta\MPl^2(M) R-\delta \alpha_Q(M) \frac{{Q^\lambda}_\lambda}{3}-\delta \alpha_2(M) R^2+\cdots\,,
\end{split}
\end{equation}
where the dots stand for higher order adiabatic contributions that decouple at large $m$, and
\begin{equation}\label{eq:deltacouplings}
\begin{split}
&\delta\rL(M)=\frac{1}{8\left(4\pi\right)^2}\left(M^4-4m^2M^2+3m^4-2m^4 \ln \frac{m^2}{M^2}\right),\\
&\delta\MPl^2(M) =\frac{\left(\xi-\frac{1}{6}\right)}{(4\pi)^2}\left(M^2-m^2+m^2\ln \frac{m^2}{M^2}\right),\\
&\delta \alpha_Q(M)=-\frac{1}{2(4\pi)^2}\ln\frac{m^2}{M^2},\\
&\delta{\alpha_2}(M)=\frac{\left(\xi-\frac{1}{6}\right)^2}{4(4\pi)^2}\ln\frac{m^2}{M^2}.
\end{split}
\end{equation}
 These quantities  are finite renormalization effects which are generated in the  subtraction \eqref{eq:LWrenormalized}. We have defined $\delta\rL(M)=\rL(M)-\rL(m)$ and $\delta\MPl^2(M)= \MPl^2(M)-\MPl^2(m)$. In general, for two arbitrary values $M_1$ and $M_2$ of the scale, we have
\begin{equation}\label{eq:rLdif}
\rho_\Lambda(M_2)-\rho_\Lambda(M_1 )=\frac{1}{8(4\pi)^2}\left(M_2^4-M_1^4-4m^2(M_2^2-M_1^2)-2m^4\ln \frac{M_1^2}{M_2^2}\right)\,,
\end{equation}
\begin{equation}\label{eq:MPLdif}
\MPl^2(M_2)-\MPl^2(M_1)=\frac{\left(\xi-\frac{1}{6}\right)}{(4\pi)^2}\,\left(M_2^2-M_1^2+m^2\ln \frac{M_1^2}{M_2^2}\right)\,.
\end{equation}
Similarly with the dimensionless coefficients $\alpha_2$ and $\alpha_Q$ of the HD terms.

We confirm from the above results that the dependence on  $\mu$ has canceled along with the poles. It should be emphasized that the subtracted term $L_{\rm div}(M)$ at the scale $M$  in \eqref{eq:LWrenormalized} involves not just the UV-divergences but also the full expression obtained up to adiabatic order 4 ($j=0,1,2$)  in the DeWitt-Schwinger expansion\,\eqref{eq:effLagrangian}, and therefore it includes their finite parts. It follows that no arbitrary additive constants are left in the subtraction. This is consistent with the off-shell ARP procedure \eqref{RenormalizedEMTScalar} for the EMT.
We conclude that the renormalized effective action of vacuuum reads\footnote{The vacuum effective action  depends on the renormalization scale $M$ since it is only a part of the full effective action. In fact, in the QFT context the classical part of the action, Eq.\eqref{eq:LEHHD}, is also dependent on $M$ through the running couplings. This is how the full renormalized effective action is independent of $M$, as the bare action itself.}
\begin{equation}\label{eq:effActionLren}
\begin{split}
W_{\rm ren}(M)=\int d^4 x\sqrt{-g} \left( \delta \rL(M)-\frac{1}{2}\delta\MPl^2(M) R-\delta \alpha_Q(M) \frac{{Q^\lambda}_\lambda}{3}-\delta \alpha_2(M) R^2\right)\,.
\end{split}
\end{equation}
In turn the renormalized vacuum EMT now follows from computing the functional derivative in  Eq.\eqref{eq:DefWeff} using the above $W_{\rm ren}(M)$.
Notice that we may drop  the contribution  from the Euler density  $E$ in ${Q^\lambda}_\lambda$, Eq.\,\eqref{eq:traceQ1}, since the functional variation of the Gauss-Bonnet term associated with it is exactly zero in $n=4$ spacetime dimensions. Similarly we may drop also the total derivative term $\Box R$ at the level of the action.  The final result for the  renormalized vacuum EMT is the following:

\begin{equation}\label{eq:TrenMm}
\begin{split}
\langle T_{\mu\nu}^{\delta \phi}\rangle_{\rm ren}(M)= \delta\MPl^2(M) G_{\mu\nu}+ \delta \rL(M) g_{\mu\nu}+\delta\alpha(M)\leftidx{^{(1)}}{\!H}_{\mu\nu}-\frac{1}{30}\delta\alpha_Q(M)\left(\leftidx{^{(2)}}{\!H}_{\mu\nu}-\frac13 \leftidx{^{(1)}}{\!H}_{\mu\nu}\right)\,,
 \end{split}
\end{equation}

where we have used $2\delta\alpha_2=\delta\alpha$ and we recall that the coefficients of the various tensor terms on the \textit{rhs} of the previous formula are given explicitly  by equations\,\eqref{eq:deltacouplings}.
The last three terms in that formula involve the standard covariantly conserved HD tensors  $\leftidx{^{(1)}}{\!H}_{\mu\nu}$ and $\leftidx{^{(2)}}{\!H}_{\mu\nu}$, which are essentially given by the metric functional derivatives of $R^2$ and $R_{\mu\nu}R^{\mu\nu}$, respectively\,\cite{BirrellDavies82}. For the FLRW spacetime these  two HD tensors are not independent, they are related as $\leftidx{^{(2)}}{\!H}_{\mu\nu}=\frac13 \leftidx{^{(1)}}{\!H}_{\mu\nu}$, which is a direct consequence of the vanishing of the Weyl tensor for conformally flat backgrounds. Thus,  the previous equation boils down to
\begin{equation}\label{eq:TrenMm0FLRWMm}
\begin{split}
\langle T_{\mu\nu}^{\delta \phi}\rangle_{\rm ren}(M)=
& \delta \rL(M) g_{\mu\nu}+\delta\MPl^2(M) G_{\mu\nu}+\delta\alpha(M)\leftidx{^{(1)}}{\!H}_{\mu\nu}\,.
 \end{split}
\end{equation}
Finally, the ZPE is the $00th$-component of the vacuum EMT. Expressing the result in terms of the cosmic time and the corresponding Hubble function $H=\dot{a}/a$, we find the renormalized ZPE:
\begin{equation}\label{eq:T00Integrated}
\begin{split}
&\left\langle T_{00}^{\delta \phi}\right\rangle_{\rm ren}(M)
=\frac{a^2}{128\pi^2}\left(-M^4+4m^2M^2-3m^4+2m^4\ln \frac{m^2}{M^2}\right)\\
&- \left(\xi-\frac{1}{6}\right)\frac{3 a^2 H^2}{16\pi^2}\left(m^2-M^2-m^2\ln \frac{m^2}{M^2}\right)\\
&+\left(\xi-\frac{1}{6}\right)^2\frac{9a^2}{16\pi^2}\left(6H^2\dot{H}+2H\ddot{H}-\dot{H}^2\right)\ln \frac{m^2}{M^2}+\mathcal{O}\left(\frac{H^6}{m^2}\right),
\end{split}
\end{equation}
where we have used the  expressions for $G_{00}=3a^2H^2$ and $\leftidx{^{(1)}}{\!H}_{00}= -18 a^2\left(\dot{H}^2-2H\ddot{H}-6H^2\dot{H}\right)$ in the flat FLRW metric and the coefficients \eqref{eq:deltacouplings} of the effective Lagrangian. The notation $\mathcal{O}(H^6 / m^2)$ schematically denotes the terms of adiabatic order 6 (consisting of 6 time derivatives of the scale factor in different combinations, many of them involving time derivatives of $H$), which will not be addressed here. The interested reader can check \cite{CristianJoan2022a} for more details.

It is important to realize at this point that the full vacuum EMT must include the (renormalized) $\rL$ term in the EH action \eqref{eq:EH}, that is to say,
\begin{equation}\label{RenEMTvacuum}
\langle T_{\mu\nu}^{\rm vac}\rangle_{\rm ren}(M)=-\rho_\Lambda (M) g_{\mu \nu}+\langle T_{\mu \nu}^{\delta \phi}\rangle_{\rm ren}(M)\,.
\end{equation}
This equation is also valid for the bare values, of course.
The VED as measured by an observer with 4-velocity $U^\mu$ can now be extracted from \eqref{RenEMTvacuum} as follows: $\rv=\langle  T_{\mu\nu}^{\rm vac} U^\mu U^\nu\rangle$.
In the rest frame of the observer, we have $U^\mu=\left(1/\sqrt{-g_{00}},0,0,0\right)=\left(1/a,0,0,0\right)$, which correctly satisfies $g_{\mu\nu}U^\mu U^\nu=-1$. Thus,  the renormalized VED reads
\begin{equation}\label{RenVDE}
\rv(M)= \frac{\langle T_{00}^{\rm vac}\rangle_{\rm ren}(M)}{a^2}=\rho_\Lambda (M)+\frac{\langle T_{00}^{\delta \phi}\rangle_{\rm ren}(M)}{a^2}\,.
\end{equation}
Insofar as concerns the vacuum pressure, we should {\em not} pretend a priori that its equation of state (EoS) is $\Pv(M)=-\rv(M)$ since quantum effects can modify it. However, we may licitly assume that the vacuum behaves as a perfect fluid, $\langle \left(T^{\rm vac}\right)^\mu_{\,\nu}\rangle=\Pv\,\delta^\mu_{\,\nu}+\left(\rv+\Pv\right)U^\mu U_\nu$. This assumption is perfectly consistent with Eq.\,\eqref{RenVDE}.
In fact, by just rewriting \eqref{RenEMTvacuum} in the alternative form $\langle \left(T^{\rm vac}\right)^\mu_{\,\nu}\rangle=-\rho_\Lambda \delta^\mu_{\,\nu}+g^{\mu\alpha}\langle T^{\delta \phi}_{\alpha\nu}\rangle$ we may combine the last two equations and find the expression for the  VED  as
$\rv=-\langle \left(T^{\rm vac}\right)^0_{\, 0}\rangle=\rL-g^{00}\langle T_{00}^{\delta \phi}\rangle=\rho_\Lambda+\frac{\langle T_{00}^{\delta \phi}\rangle}{a^2}$,
which is just Eq.\eqref{RenVDE}.

By the same token, upon taking the trace of the above perfect fluid equation we find the vacuum pressure: $\Pv=\frac13 \left(\rv +\langle T^{\rm vac}\rangle\right)$, where $ T^{\rm vac}\equiv g^{\mu\nu}\, T^{\rm vac}_{\mu\nu}$ is such a trace. The latter can also be computed by tracing over Eq.\,\eqref{RenEMTvacuum}, which yields  $\langle T^{\rm vac}\rangle=-4\rL+\langle T^{\delta \phi}\rangle$, with $T^{\delta \phi}\equiv g^{\mu\nu}\, T^{\delta \phi}_{\mu\nu}$. On combining them both and using Eq.\,\eqref{RenVDE} to eliminate $\rL$ in favor of $\rv$, we find the final sought-for equation for the vacuum pressure. In terms of the renormalized quantities at the scale $M$, it reads
\begin{equation}\label{eq:VacuumPressureDef2}
\Pv(M)=-\rv(M)+\frac{1}{3}\left( \langle T^{\delta \phi} \rangle_{\rm ren}(M)+4\frac{\langle T_{00}^{\delta \phi} \rangle_{\rm ren}(M)}{a^2}\right)\,.
\end{equation}
Incidentally, one can easily show that this equation can also be obtained from $\Pv=\langle T_{ii}^{\delta \phi}\rangle_{\rm ren}(M)/{a^2}$ (which follows from the perfect fluid form of the vacuum EMT, without the sum over $i$, for any $i=1,2,3$) using the isotropy condition.  It is apparent from the result \eqref{eq:VacuumPressureDef2} that the effective EoS of vacuum deviates from the naive one $\Pv=-\rv$ by terms which depend on the VEV of the trace of the  EMT as well as on the ZPE. These terms obviously stand both for quantum corrections to the classical EoS of vacuum.

Notice that $\langle T_{00}^{\delta \phi} \rangle_{\rm ren}(M)$ on the \textit{rhs} of Eq.\,\eqref{eq:VacuumPressureDef2} has already been computed in \eqref{eq:T00Integrated}.  However, in order to get an explicit expression for the vacuum pressure, we need to compute also the VEV of the trace of the EMT. It can be found by taking the trace of the renormalized vacuum EMT, Eq.\,\eqref{eq:TrenMm0FLRWMm}. The result up to fourth adiabatic order is as follows:
\begin{equation}\label{eq:TraceIntegrated}
\begin{split}
&\left\langle T^{\delta \phi}\right\rangle_{\rm ren}(M)
=\frac{1}{32\pi^2}\left(3m^4-4m^2M^2+M^4-2m^2\ln \frac{m^2}{M^2}\right)\\
&+\left(\xi-\frac{1}{6}\right)\frac{3}{8\pi^2} \left(2H^2+\dot{H}\right) \left(m^2-M^2-m^2\ln \frac{m^2}{M^2}\right)\\
&-\left(\xi-\frac{1}{6}\right)^2\frac{9}{8\pi^2}\left(12H^2 \dot{H}+4\dot{H}^2+7H\ddot{H}+\vardot{3}{H}\right)\ln \frac{m^2}{M^2}+\mathcal{O}\left(\frac{H^6}{m^2}\right)\,.
\end{split}
\end{equation}
Substituting \eqref{eq:T00Integrated} and \eqref{eq:TraceIntegrated} into \eqref{eq:VacuumPressureDef2} and after some rearrangements we find that the vacuum pressure can be written in a remarkably simple way:
\begin{equation}\label{eq:fullpressure}
\begin{split}
\Pv(M)=&-\rv(M)+\left(\xi-\frac{1}{6}\right)\frac{1}{8\pi^2}\dot{H}\left(m^2-M^2-m^2\ln\frac{m^2}{M^2}\right)\\
&-\frac{3}{8\pi^2}\left(\xi-\frac{1}{6}\right)^2\left(6\dot{H}^2+3H\ddot{H}+\vardot{3}{H}\right)\ln \frac{m^2}{M^2}\,.
\end{split}
\end{equation}
In particular, we note that all the quartic mass terms $\sim m^4$ introduced by the quantum corrections on the \textit{rhs} of Eq.\,\eqref{eq:VacuumPressureDef2} have  exactly canceled out in the final expression for the vacuum pressure. In addition, we find that the net quantum correction to the classical vacuum EoS  vanishes identically for $H=$const. since it depends only on time derivatives of $H$.  This feature will play an important role when we discuss the inflationary epoch.  We should also remark that the mentioned quantum corrections also vanish identically in the conformal limit, $\xi=1/6$, as could be expected.

\subsection{Running Vacuum and EoS in the Late Universe}\label{sec: Eosvac}

In the current universe the ${\cal O}(H^4)$ terms are negligible and hence from equations \eqref{eq:T00Integrated} and \eqref{RenVDE} we find the leading form of the VED at low energy:
\begin{equation}\label{RenVDEexplicit}
\begin{split}
\rv(M,H)&= \rho_\Lambda (M)+\frac{1}{128\pi^2 }\left(-M^4+4m^2M^2-3m^4+2m^4 \ln \frac{m^2}{M^2}\right)\\
&+\left(\xi-\frac{1}{6}\right)\frac{3 {H}^2 }{16 \pi^2 }\left(M^2-m^2+m^2\ln \frac{m^2}{M^2} \right)\,,
\end{split}
\end{equation}
and from \eqref{eq:fullpressure} we find the leading expression for the pressure:
\begin{equation}\label{eq:leadingpressure}
\begin{split}
\Pv(M,H)=&-\rv(M,H)+\left(\xi-\frac{1}{6}\right)\frac{1}{8\pi^2}\dot{H}\left(m^2-M^2-m^2\ln\frac{m^2}{M^2}\right)\,.
\end{split}
\end{equation}
Let us now focus for a while on the VED \eqref{RenVDEexplicit}. As is mandatory in any calculation in QFT, a renormalization scale $M$ appears in the result. For a physical interpretation,  an appropriate choice of that scale is necessary when we focus on just a part of the full action, in this case the vacuum effective action. To deal with this fact, let us first note that in the above expressions, the values of $M$ and $H$ are independent, so we may compute the difference between the VED values at points $(M,H)$ and $(M_0,H_0)$.  The result is the following:
\begin{equation}\label{DiffHH0MM0}
\begin{split}
\rv(M,H)-\rv(M_0,H_0)&
=\frac{3\left(\xi-\frac{1}{6}\right)}{16\pi^2}\left[H^2\left(M^2-m^2+m^2\ln\frac{m^2}{M^2}\right)\right.\\
&\left.-H_0^2\left(M_0^2-m^2+m^2\ln\frac{m^2}{M_0^2}\right)\right]\,,
\end{split}
\end{equation}
where we should note that the quartic mass terms have canceled out upon using Eq.\eqref{eq:rLdif}. For a physical interpretation,  it is reasonable that $M$ should be near the characteristic energy scale of the FLRW spacetime at the given cosmic epoch. It has been proposed that the value of the Hubble rate $H$  satisfies this condition because this choice should have physical significance in the cosmological context\,\cite{CristianJoan2020,CristianJoan2022a}. Therefore,  we set $M=H$ and $M_0=H_0$ and we can neglect the $\mathcal{O}(H^4)$ for the present universe. The outcome is

\begin{equation}\label{DiffVEDphys}
\begin{split}
\rv(H)-\rv(H_0)&=\frac{3\left(\xi-\frac{1}{6}\right)}{16\pi^2}\left[H^2\left(H^2-m^2+m^2\ln\frac{m^2}{H^2}\right)-H_0^2\left(H_0^2-m^2+m^2\ln\frac{m^2}{H_0^2}\right)\right]+\cdots\\
\simeq& \frac{3\left(\xi-\frac16\right)m^2 }{16\pi^2}\left[-\left(H^2-H_0^2\right)+H^2\ln\frac{m^2}{H^2}-H_0^2\ln\frac{m^2}{H_0^2}\right]\,,
\end{split}
\end{equation}

where we have defined  $\rho_{\rm vac}(H)\equiv\rho_{\rm vac}(M=H,H)$ and similarly $\rho_{\rm vac}(H_0)\equiv\rho_{\rm vac}(H_0,H_0)$.
This equation provides the VED at the scale $M=H$ in terms of the VED at another renormalization scale $M_0=H_0$, and hence it  expresses the `running' of the VED between the two scales. One can see that such running is manifestly  slow in the current universe, which is consistent with the fact that the description of the cosmological expansion with a rigid  CC is approximately possible (as it is the case of the $\CC$CDM).  However, the present RVM picture predicts a small departure from that rigid behavior since the VED is evolving slowly with the expansion, and hence the effective CC too. As it is apparent, for the minimal coupling situation ($\xi=0$) there is still running of the VED, and this is also true for any non-minimal coupling value  $\xi$ of the scalar field with gravity, except for the case $\xi=1/6$, where there is no running at all. This was expected because of conformal invariance.  In addition, owing to the aforementioned absence of $\sim m^4$ contributions in this framework (which can be extremely large for any typical particle in the standard model of particle physics),  there is no need to perform an unnatural fine-tuning of the VED within this renormalization context. This fact obviously alleviates the CCP in the RVM approach\cite{JSPRev2022}.

It is convenient to rewrite Eq.\,\eqref{DiffVEDphys} as follows
\begin{equation}\label{eq:RVMform}
\rv(H)\simeq \rvo+\frac{3\nu(H)}{8\pi}\,(H^2-H_0^2)\,\mpl^2\,,
\end{equation}
where $\rvo\equiv\rv(H_0)$ is identified with today's value of the VED through the measured CC.
In addition, we have introduced the effective `coefficient'
\begin{equation}\label{eq:nueff2}
\nu(H)\equiv\frac{1}{2\pi}\,\left(\xi-\frac16\right)\,\frac{m^2}{\mpl^2}\left(-1+\ln \frac{m^2}{H^{2}}-\frac{H_0^2}{H^2-H_0^2}\ln \frac{H^2}{H_0^2}\right)\,.
\end{equation}
In point of fact, $\nu(H)$ is not a numerical coefficient, since it is a function of $H$. However, due to the log behavior, it changes very slowly with the Hubble rate and its evolution from the last term quickly becomes suppressed for higher values of $H$ above $H_0$. Moreover, because $\ln \frac{m^{2}}{H^2}\gg1$ at present (this being actually true as well for the entire post-inflationary history of the universe, see Sec.\ref{sec:RVMInflation}) ,  $\nu(H)$ can be approximated by the effective parameter
\begin{equation}\label{eq:nueffAprox2}
\nueff=\frac{1}{2\pi} \left( \xi-\frac{1}{6}\right) \frac{m^2}{m_\mathrm{Pl}^2}\left(-1+\ln\frac{m^2}{H^2}\right)\simeq\epsilon\ln\frac{m^2}{H_0^2}\,,
\end{equation}
where
\begin{equation}\label{eq:epsilonparameter}
\epsilon=\frac{1}{2\pi}\,\left(\xi-\frac{1}{6}\right)\,\frac{m^2}{\mpl^2}\,.
\end{equation}
Despite that both parameters are small ($|\nueff|, |\epsilon|\ll 1$), we have $\nueff\gg\epsilon$ since $\ln\frac{m^2}{H_0^2}={\cal O}(100)$.

Therefore,  for practical purposes we can write \eqref{eq:RVMform} as follows:
\begin{equation}\label{eq:RVMcanonical}
\rv(H)=\rvo+\frac{3\nueff}{8\pi G_N}\,(H^2-H_0^2)\,.
\end{equation}
Herein $G_N\equiv 1/\mpl^2$ is assumed to be the currently  measured value of the gravitational constant.
It is remarkable that this expression for the VED turns out to adopt the canonical RVM form, see \,\cite{JSPRev2022} and references therein\footnote{We note that the canonical RVM form \eqref{eq:RVMcanonical} of the VED, which in our case emerges from off-shell ARP renormalization of the EMT in  QFT in curved spacetime, has also been independently highlighted in recent studies of dynamical dark energy in the context of lattice quantum gravity using also the same scale setting $M=H$\,\cite{Dai:2024vjc}. Remarkably enough, these authors obtain numerical lattice calculation estimates for $\nu$ in the ballpark of the fitted values for this parameter from the analyses of cosmological observations\,\cite{Sola:2015wwa,SolaPeracaula:2016qlq,SolaPeracaula:2021gxi}. For recent work exploring the running vacuum energy density in cosmology from different perspectives, see e.g. \cite{Montani:2024buy}.}.  The implied evolution of the VED is clearly moderate, as the effective parameter $\nueff$ is expected to be small due to its proportionality to $m^2 / \mpl^2\ll1$. However, the main contribution should come from scalar fields originating from a typical Grand Unified Theory (GUT) at the scale $M_X\sim 10^{16}$ GeV.  Their large masses and  large multiplicities  can make $\nueff$ still sizable for phenomenological considerations, typically one expects $\nueff\sim 10^{-3}$, although this theoretical estimate may vary within a few orders of magnitude around this value\,\cite{Fossil2008}.

From the phenomenological perspective, studies based on fitting the above RVM formula to the overall cosmological data yield an estimate for $\nueff$ at the level of  $\nueff \sim 10^{-2}-10^{-4}$\,\cite{SolaPeracaula:2021gxi}.  The order of magnitude is therefore reasonable, since it falls within the expectations of the mass spectrum and large particle multiplicities in a typical GUT. Noteworthy, too,  is the fact that these estimates on $\nueff$ lie in the ballpark  of the primordial Big Bang Nucleosynthesis (BBN) bounds obtained for this parameter in\,\cite{Asimakis:2021yct}. In actual fact the value of $\nu(H)$ around the BBN is smaller than $\nueff$ since $H$ is larger than $H_0$.

\begin{figure}[t]
  \begin{center}
    \begin{tabular}{cc}
      \resizebox{0.48\textwidth}{!}{\includegraphics{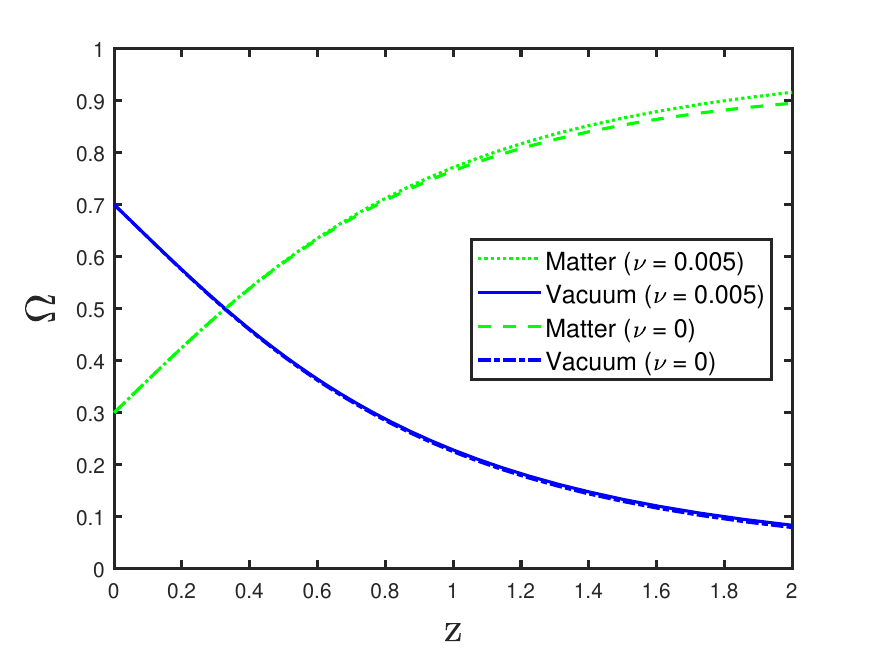}} &
      \hspace{0.3cm}
      \resizebox{0.48\textwidth}{!}{\includegraphics{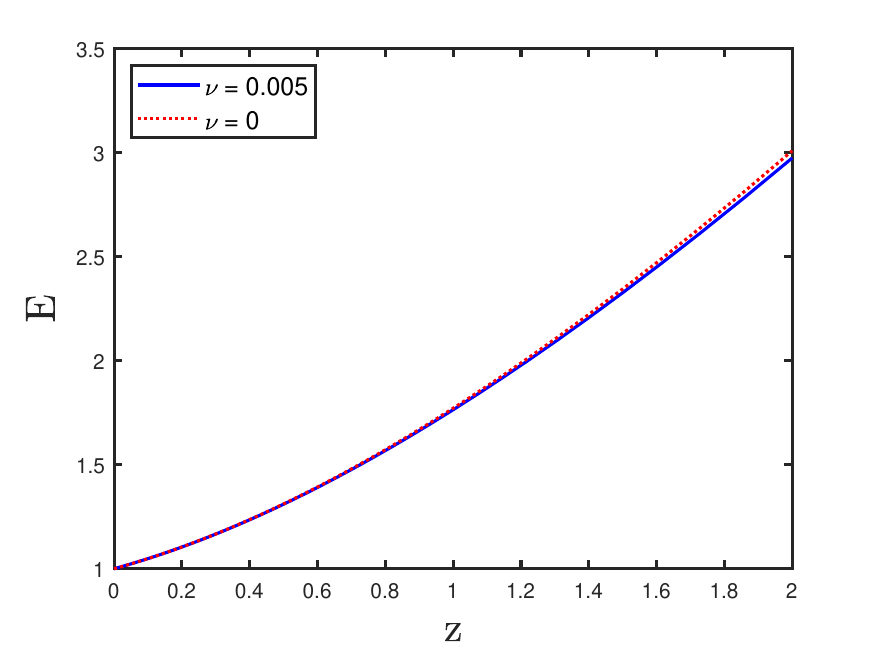}} \\
      (a) & (b)
    \end{tabular}
\caption{\textbf{(a)} Evolution of energy densities with the redshift for different values of $\nueff$; \textbf{(b)} Similarly for the normalized Hubble rate $E(z)=H(z)/H_0$. Differences are small for these observables with the typically small values of $\nueff$. The background evolution is essentially $\CC$CDM-like.}
\label{Fig:OmegasE}
  \end{center}
\end{figure}

The equation of state (EoS) analysis is always very important in studies of the DE. Using  the RVM expression for the vacuum  pressure, Eq.\,\eqref{eq:leadingpressure},  we can obtain the leading form for the effective EoS of the running vacuum:
\begin{equation}\label{eq:EoS1}
w_{\rm vac}=\frac{P_{\rm vac}(H)}{\rho_{\rm vac}(H)} \simeq-1+ \left(\xi-\frac{1}{6}\right)\frac{\dot{H} m^2}{8 \pi^2 \rv(H)}\left(1 -\ln \frac{m^2}{H^2} \right)\,.
\end{equation}
As it is obvious, $\wv$ does not remain constant with the expansion; it takes on the naive value $\wv=-1$ only for $H=$const. But this is never so in the current universe and therefore the vacuum EoS is generally dynamical in QFT. The above equation can be further worked out using equations\,\eqref{eq:RVMform}-\eqref{eq:nueff2} for the VED:
\begin{equation}\label{eq:EoS2}
w_{\rm vac}(z)=-1 +\frac{\nu_{\rm eff}\left(\Omega_{\rm m}^0 (1+z)^3+\frac{4}{3}\Omega_{\rm r}^0 (1+z)^4\right)}{\Omega_{\rm vac}^0+\nu_{\rm eff}\left(-1+\Omega_{\rm m}^0 (1+z)^3 +\Omega_{\rm r}^0(1+z)^4+\Omega_{\rm vac}^0 \right)}\,,
\end{equation}
where we have used the definition of $\nueff$ given in \eqref{eq:nueffAprox2} and the current cosmological parameters for matter and radiation: $\Omega^0_i=\rho^0_i/\rho^0_c=8\pi G_N\rho^0_i/(3H_0^2)$. Remarkably, the EoS for the cosmological vacuum is a function of the redshift. The term producing a deviation from the traditional value $w_{\rm vac}=-1$ is proportional to $\nueff$ and involves the leading quantum effects. Thus $\nueff$ is responsible both for the dynamics of the VED and for the dynamics of its EoS.
For very low redshift  $z$ the above expression is easily seen to boil down to the very simple form
\begin{equation}\label{EqStateScalar}
w_{\rm vac}(z)\simeq -1+\nu_{\rm eff}\frac{\Omega_{\rm m}^0}{\Omega_{\rm vac}^0}(1+z)^3.
\end{equation}
%
\begin{figure}[t]
  \begin{center}
    \begin{tabular}{cc}
      \resizebox{0.50\textwidth}{!}{\includegraphics{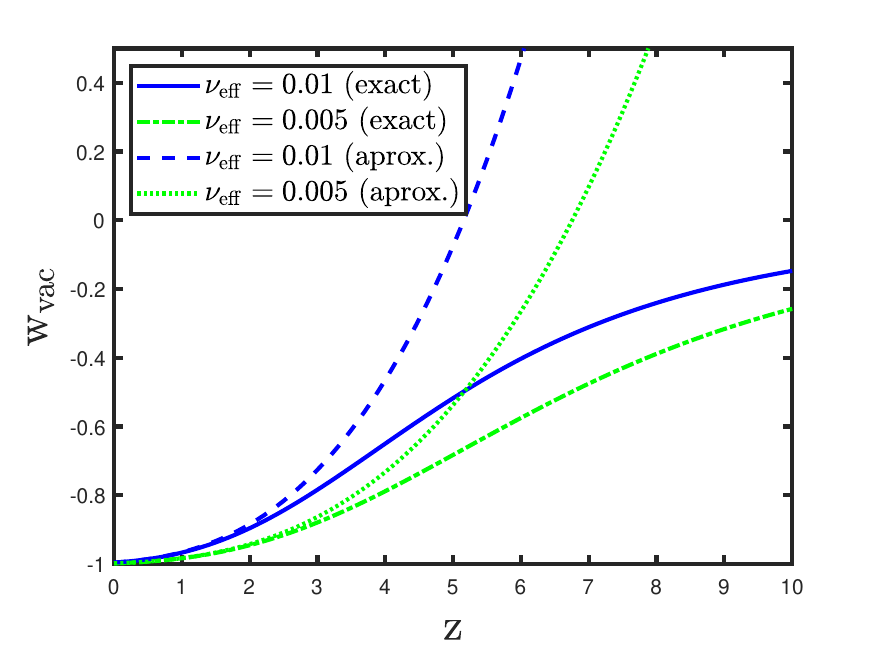}} &
      \hspace{0.0cm}
      \resizebox{0.50\textwidth}{!}{\includegraphics{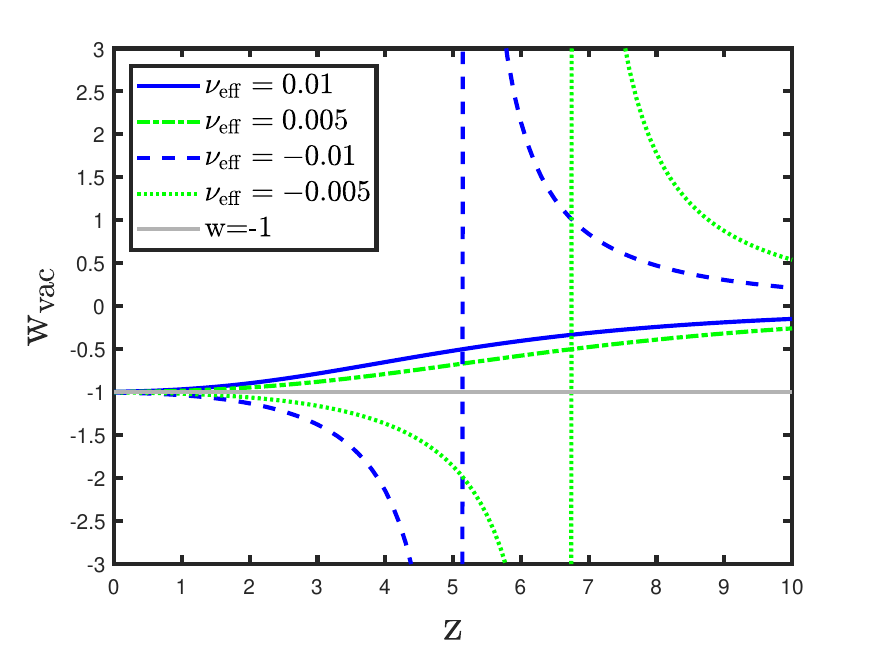}} \\
      (a) & (b)
    \end{tabular}
\caption{\textbf{(a)} EoS of the running vacuum $\wv(z)$  evolving with the redshift in the late universe. It is shown both the exact \eqref{eq:EoS2} and the approximate \eqref{EqStateScalar} formulas for different positive values of $\nueff$. The approximate curves  deviate significantly from the exact ones for $z>3$; \textbf{(b)} Here we consider both signs of $\nueff$, and all curves follow the exact formula \eqref{eq:EoS2}. The vertical asymptotes appear only for $\nueff<0$, see the text for an explanation. }
\label{Fig:EosvacLateUniv}
  \end{center}
\end{figure}
The following interesting conclusion ensues. If we assume a positive sign for $\nueff$,  the above equation predicts that the vacuum energy behaves as quintessence ($w_{\rm vac}(z)\gtrsim-1$) around the current time, whereas if $\nueff<0$ it behaves effectively as phantom DE ($w_{\rm vac}(z)\lesssim-1$). It is interesting to note that for the typical values of $\nueff$ mentioned above,  the small deviation from -1 could be subject of measurement around the present time and might act as a smoking gun of the underlying RVM mechanism.  Although the EoS formula \eqref{EqStateScalar} is valid only for small values of the redshift $z$, the more precise formula \eqref{eq:EoS2} shows that at very high redshifts $\wv\to 0$ or $\wv\to 1/3$ depending on whether we are in the matter- or radiation-dominated epochs, which are controlled by the redshift factors $(1+z)^3$ and $(1+z)^4$ respectively. Thus, the vacuum adopts a kind of `chameleonic' behavior: it tracks the EoS of matter at different epochs; see \cite{CristianJoan2022b} for more details. All in all, we should emphasize that these are dynamical properties of the quantum vacuum which we have obtained from first principles, namely from explicit QFT calculations in the FLRW background.
\begin{figure}[t]
  \begin{center}
      \resizebox{0.68\textwidth}{!}{\includegraphics{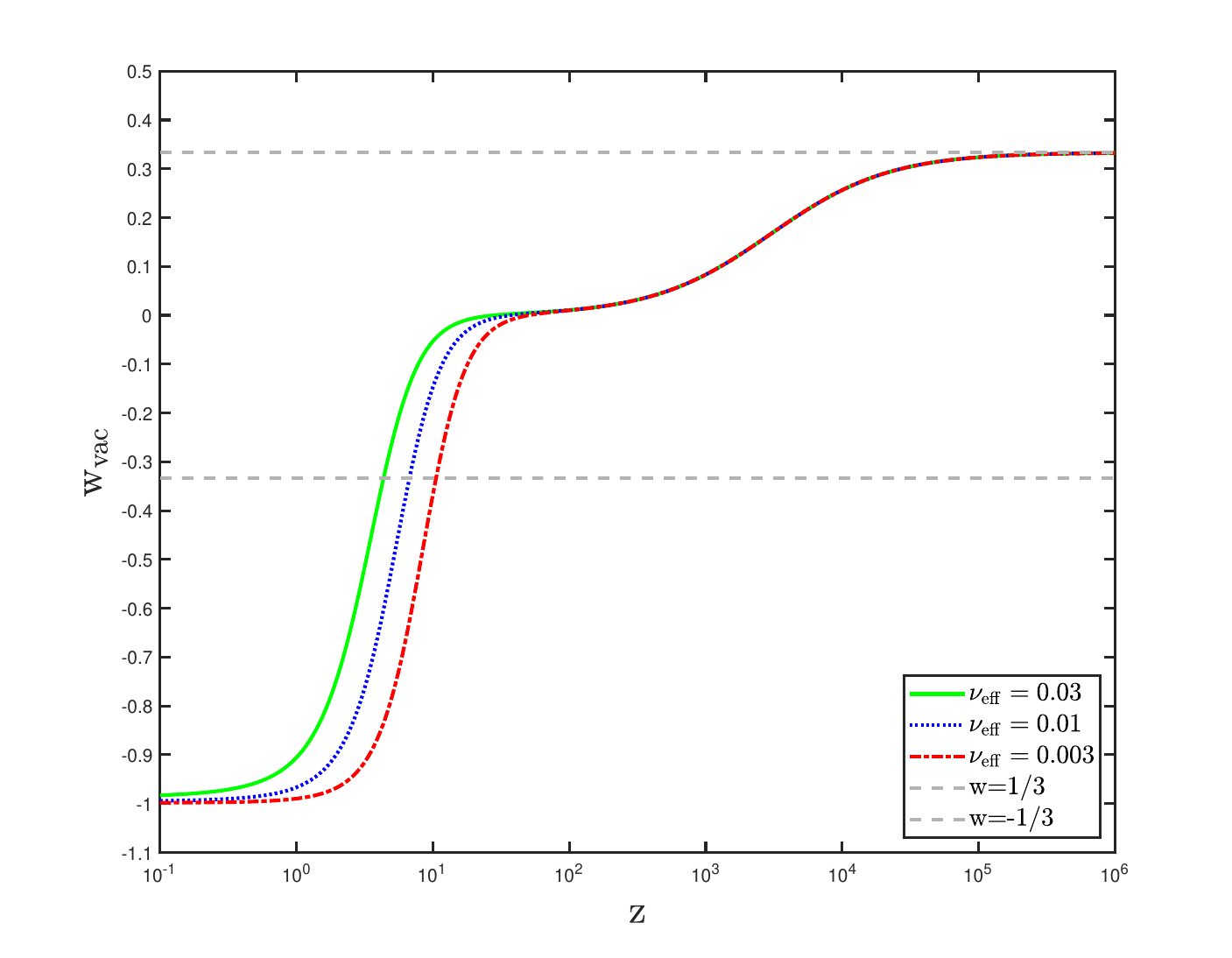}}
\caption{EoS of the running vacuum $\wv(z)$  up to a redshift range covering the entire FLRW regime for  three positive values of $\nueff$. The low  $z$ region is described in more detail in Fig. \ref{Fig:EosvacLateUniv}. The horizontal dashed lines mark the EoS value of radiation (the higher line) and the limiting value for producing acceleration (the lower one). }
\label{Fig:EoSvacExtended}
  \end{center}
\end{figure}

To graphically illustrate the above results, in Fig. \ref{Fig:OmegasE} we display the evolution of the matter and vacuum energy densities normalized with respect to the critical density, i.e. $\Omega_i(z)=\rho_i(z)/\rho_c(z)$ ($i=m,{\rm vac}$) for different values of $\nueff>0$, as well as the evolution of the normalized Hubble rate to the present value, $E(z)=H(z)/H_0$.  As could be expected, the departure from the standard model scenario is minimal for these observables and small values of $\nueff$. The truly remarkable effects deviating from the standard behavior can instead be grasped in Fig.\,\ref{Fig:EosvacLateUniv}, where we illustrate the dynamics of the vacuum EoS in the low redshift region (which comprises the part that is relevant for type Ia supernova measurements, up to $z\sim 2$), as well as in the intermediate redshift region. In Fig.\ref{Fig:EosvacLateUniv}(a) we assume $\nueff>0$ and include two sets of curves, one set (with dashed lines) based on the approximate EoS \eqref{EqStateScalar} and another set (with solid lines)  based on the more accurate expression \eqref{eq:EoS2}. For relatively low redshifts ($z<3$), the approximate EoS formula works pretty well, but at higher redshifts the more precise form becomes necessary. In Fig.\ref{Fig:EosvacLateUniv}(b), instead,  all curves use the exact expression \eqref{eq:EoS2}. In it, we superimpose curves with both signs of $\nueff$. We can see that for $\nueff<0$ vertical asymptotes appear, caused by the fact that the denominator of Eq. \eqref{eq:EoS2} vanishes. However, this is not a real singularity since the energy densities remain finite at all times, see e.g. \cite{Staro} and \cite{SolaStefancic}.  Finally, in Fig. \ref{Fig:EoSvacExtended} we display the EoS of the running vacuum in the extended redshift domain comprising the matter-dominated epoch up to the radiation-dominated epoch, where one can appreciate very clearly that at higher and higher redshifts the vacuum EoS tends to adopt the EoS of the dominant matter component (namely, first $\wv\simeq 0$ and subsequently $\wv\simeq 1/3$).  This curious `chameleonic nature' of the quantum vacuum, referred to before, is caused by the quantum effects on the classical description.  Despite the fact that $\wv\simeq -1$ at present, the quantum corrections induce a deviation of the standard expectations, which leads to an effective quintessence or phantom-like behavior (depending on the sign of $\nueff$). The canonical form $w=-1$ is recovered again, with great precision, in the very early universe where inflation occurs. Nevertheless, we cannot show this part in Fig.\,\ref{Fig:EoSvacExtended} since the VED must first incorporate additional quantum effects that we have neglected in the low-energy domain.  We treat this situation in detail in the next section.

On the phenomenological side, the situation concerning the possible detection of signs of dynamical DE looks encouraging in view of the recent data releases from the DESI collaboration, see \cite{DESI1,DESI2,DESI3}. The results using various parameterizations of the DE seem to point to quintessence behavior at low redshifts. This would be consistent with $\nueff>0$ within the RVM.  However the EoS behavior at higher redshifts is more difficult to assess using simple parameterizations. In fact, direct comparison between parameterizations and models in wide ranges of redshif must be done with caution. We will have to wait until more data will be collected in the future.

\section{$H^4$-Inflation from Running Vacuum}\label{sec:RVMInflation}

Inflation is a cosmic phenomenon that is necessary to cure a variety of serious inconsistencies of the $\CC$CDM. In its absence, we could not understand how causality laws could arrange for securing the homogeneity and isotropy of the observed CMB, for instance, or how to explain the high level of spatial flatness of the universe at present without fine-tuning, or even its large amount of entropy today: $S\sim 10^{88}$ (in natural units).  Since the $\CC$CDM does not have the capacity to solve any of these problems and related ones, which originate deep in the remote past, they are usually `fixed' by postulating the existence of a devoted scalar field (called the ``inflaton'') which takes care of these arrangements during the very first stages of the cosmic history\,\cite{KolbTurner}.  Even if this quick fix or patchup of the malfunctioning of the $\CC$CDM in the early stages of the universe can be efficiently implemented and is generally considered acceptable, we must recognize that it is not very natural, since it is ultimately \textit{ad hoc}. It should be much more natural that a unified theory of the cosmic evolution could explain the correct expansion history from end to end.

From our former considerations we have seen that the structure of the RVM contains not only the necessary ingredients to influence on the physics of the current epoch through a mildly time-evolving VED, but also to account for the evolution in the  opposite end of the cosmic span, namely at an epoch where the higher powers of the Hubble rate can be really significant, in fact dominant. Thus, remarkably enough, this crucial segment of the very early cosmic history also comes about encoded in the RVM framework. Indeed, once the vacuum energy density in cosmological spacetime is renormalized through the off-shell adiabatic procedure a definite prediction for a new mechanism of fast inflation naturally emerges, which is characterized by a short stage where $H\simeq H_I=$const. early on in the cosmic history.  The  value $H_I$ must, of course,  be very large, presumably around a characteristic GUT scale just below the Planck mass.  The quantized matter fields in the GUT constitute the material support for this framework.  Hence no \textit{ad hoc} scalar field  potential associated to some inflaton is required\,\cite{KolbTurner}. It is instead the pure work of gravity fueled by the quantum matter effects of the semiclassical theory.  Such an alternative form of  inflation,  based on the constancy of $H$ for a short lapse of cosmic time,   is what we have called  `RVM-inflation'.  To bring about an inflationary period with this mechanism, we need powers of $H$ higher than  $H^2$. Since only even powers of the Hubble rate are permitted by the general covariance of the effective action, the next leading power is $H^4$. These powers are in fact available in our result \eqref{DiffHH0MM0} when we implement our setting $M=H$ for the renormalization scale, in the same way as we did for the current universe.  Given the fact that $H>m$ for the very early universe ($m$ being a typical particle mass in the GUT), and that $H_0\ll m$ for the current Hubble parameter, we can neglect the contribution from the VED at present. As a result, within a very good approximation the VED at the early universe can be expressed as follows:
\begin{equation}\label{eq:VEDinfl}
\rho_\mathrm{vac}(H)\simeq\frac{3\left( \xi-\frac{1}{6}\right)}{16\pi^2}\left[ H^4+H^2m^2\left(\ln\frac{m^2}{H^2}-1\right)\right]\,,
\end{equation}
where for simplicity we assume that the scalar fields of the GUT dominate the VED and that their effect is   represented on average by a single component of mass $m$. It is understood that multiplicity factors may account for additional contributions.
In the above equation for the VED, the subleading effect attached to the $H^2$ term is still kept as a correction, but now the obvious leading power for the inflationary phase  is $H^4$.  The above equation can be conveniently rewritten as follows:
\begin{equation}\label{eq:VEDinf2l}
\rho_\mathrm{vac}(H)=\frac{3\left( \xi-\frac{1}{6}\right)}{16\pi^2}\, H^4+\frac{3{\nu}(H)}{8\pi G_N}\,H^2\,.
\end{equation}
In the above expression we can use
\begin{equation}\label{eq:nuH}
{\nu}(H)\simeq\epsilon \left(-1+\ln\frac{m^2}{H^2}\right)\,,
\end{equation}
with $\epsilon$ as in \eqref{eq:epsilonparameter}. Notice that the approximation \eqref{eq:nuH} for  $\nu(H)$ is fully justified in the high energy regime since now $H\gg H_0$ and the last term of \eqref{eq:nueff2}, which is proportional to $H_0^2$, can be neglected.   As we shall see, during the inflationary period $H$ remains essentially constant around a large value $H_I\gg m$. It follows that in this stage we can approximate $\nu(H)$ by ${\nu}(H_I)\equiv\nu_I$, given by
\begin{equation}\label{eq:nuI}
\nu_I\equiv\epsilon \ln\frac{m^2}{H_I^2}\,.
\end{equation}
This parameter is the analogue of $\nueff$ in the inflationary epoch.  We note, however,  that  $|\nu_I|\ll|\nueff|$ since the logarithmic term in \eqref{eq:nuI} is not as large (in absolute value)  as in \eqref{eq:nueffAprox2}. Below we shall be more precise about the value of $H_I$. Once inflation is left behind the more accurate form \eqref{eq:nuH} may be necessary since $H$ evolves towards values much smaller than $H_I$, so we must check its effect.

\subsection{Analytical and Numerical Solution of the $H^4$ Inflationary Scenario}\label{sec:solutions}

As a first approach in our way to  solving  the cosmological equations with the VED given by  Eq.\,\eqref{eq:VEDinf2l},  let us assume that ${\nu}(H)$ in that equation can be approximated by the mentioned constant value $\nu_I$ from \eqref{eq:nuI}. This is tantamount to assuming that the effect of the logarithmic term in it is negligible even after inflation.  We will come back to this point later on, but let us make things simple to start with.  For ${\nu}(H)=\nu(H_I)\equiv\nu_I$ constant,  the cosmological equations can be solved analytically.  In this case, equation \eqref{eq:VEDinf2l} can be considered as a particular case of the generalized VED form\footnote{In general, we could also have $G=G(H)$\ rather than just $G_N$\,\cite{CristianJoan2022a}. However, for the sake of simplicity and aiming to obtain an exact analytical solution of the cosmological equations we shall not consider this possibility here.}
\begin{equation}\label{eq:rvm1}
\rv(H)=\frac{3}{\kappa^2} \left( c_0+\nu H^2+ \frac{H^4}{H_I^2} \right)\,.
\end{equation}
We have used the notation $\kappa^2=8\pi G_N=8\pi/\mpl^2$. The above template for the VED actually encompasses the unified description of the cosmic history in the RVM context from the very early times to our day. In the absence of quadratic and quartic powers of $H$, the parameter $c_0$ in this equation would be related to the cosmological constant that we have measured today simply as $\CC=3c_0$. However, in the presence of the dynamical terms, this is no longer  true as the physical value of the cosmological term at present ($H=H_0$) is $\CC=\kappa^2\rv(H_0)$. Since $\nu$ is small and $H\ll H_I$ today, the previous relation between $\CC$ and $c_0$ is still approximately true, but the fact that it is not exactly true means that in general $\rv(H)$ evolves with the expansion, and so does $\Lambda(H)=\kappa^2\rv(H)$ in this class of models.  Near our time we can neglect the $H^4$ term, as we did previously and we recover the simpler form \eqref{eq:RVMcanonical}, whereas in the early universe the $H^4$ power becomes dominant and we may completely neglect the constant term $c_0$, which as we have seen is of the order of the measured CC value today, but we can still keep the $H^2$ term as a subdominant correction.
The cosmological equations to solve are therefore the following:
\begin{equation}
\begin{split}\label{eq:friedmann}
3H^2&=\kappa^2 (\rho_\mathrm{rad}+\rho_\mathrm{vac}) \; , \\
3H^2+2\dot{H}&=-\kappa^2 (P_\mathrm{vac}+\frac{1}{3}\rho_\mathrm{rad}) \; .
\end{split}
\end{equation}
In these equations, we are considering that at this stage of the primeval cosmic history we just have vacuum energy and relativistic particles (i.e. radiation, with EoS $w_{\rm rad}=1/3$) exchanging energy.
Notice that since we aim at an inflationary solution for $H=$const.  the vacuum pressure  satisfies the traditional equation of state $\Pv=-\rv$ during the very short inflationary period, as can be seen from Eq.\eqref{eq:fullpressure}. This allows us to combine the above two equations \eqref{eq:friedmann}, together with Eq.\,\eqref{eq:rvm1}, into the following one:
\begin{equation}\label{eq:Hdiffeq}
\dot H+2H^2=2(\nu H^2+\frac{H^4}{H_I^2})\,,
\end{equation}
where, as noted, we neglect the small parameter $c_0$ which is relevant only at the current time. It is apparent from it that there exists an inflationary solution for $H=$const.,  specifically for $H=H_I\sqrt{1-\nu}\simeq H_I$, which is the starting point for inflation.  However, there is an evolution of the Hubble rate from this point onward, which can be computed by solving Eq.\,\eqref{eq:Hdiffeq}. An exact analytical solution is possible for $\nu=$const in terms of the scale factor (recall that $d/dt=a H d/da$), with the following outcome:
\begin{equation}\label{eq:solHubble}
H(a)=\sqrt{1-\nu} \frac{H_I}{\sqrt{1+Da^{4(1-\nu)}}}\,.
\end{equation}
Introducing this result into equations \eqref{eq:friedmann} we can also solve analytically for the energy densities of vacuum and radiation:
\begin{align}\label{solDensities}
\rho_\mathrm{vac}(a)&=\frac{3H_I^2(1-\nu)(1+\nu D a^{4(1-\nu)})}{\kappa^2\ (1+D a^{4(1-\nu)})^2}\,,\\
\rho_\mathrm{rad}(a)&=\frac{3H_I^2(1-\nu)^2D a^{4(1-\nu)}}{\kappa^2 (1+D a^{4(1-\nu)})^2}\,.
\end{align}
Here $D$ is an integration constant that is determined by the condition  $\rho_\mathrm{vac}(a_\mathrm{eq})=\rho_\mathrm{rad}(a_\mathrm{eq})$. The latter defines the equality point  $a_{\mathrm{eq}}$ where the vacuum energy density at the end of the inflationary epoch equals the radiation energy density and this defines also the early start of the radiation era.  Thus, we find  $D=\frac{1}{1-2\nu}a_\mathrm{eq}^{-4(1-\nu)}\equiv a_*^{-4(1-\nu)}$. The above cosmological solution is general for a VED of the form \eqref{eq:rvm1} and for  $c_0$  null or negligible. Previous studies considered that form on pure phenomenological grounds without establishing any fundamental connection with QFT calculations, see \cite{BLS2013,Perico:2013mna,JSPRev2015,Sola:2015csa,Yu2020}.

In our case, we have derived the VED structure \eqref{eq:rvm1} in the context of QFT in curved spacetime, and we have shown that at low energies $\nu$ in \eqref{eq:rvm1}  is effectively described by $\nueff$, given by Eq.\eqref{eq:nueffAprox2}, whereas at much earlier times (viz. those relevant for the study of inflation and its connection with the radiation-dominated era) $\nu$ is essentially given by $\nu_I$, Eq.\eqref{eq:nuI}, with some evolution \eqref{eq:nuH} during the transit into the radiation-dominated epoch, whose impact we still have to carefully assess.

It is particularly convenient to re-express the above solution in terms of the rescaled variables $\hat{a}=a/a_*$ and $\Tilde{H}_I=\sqrt{1-\nu}H_I$:
\begin{align}
H(\hat{a})&=\frac{\Tilde{H}_I}{\sqrt{1+\hat{a}^{4(1-\nu)}}}\,,\label{eq:infaprox}\\
\rho_\mathrm{vac}(\hat{a})&={\rho}_I\frac{1+\nu\hat{a}^{4(1-\nu)}}{[1+\hat{a}^{4(1-\nu)}]^2}\label{eq:infaprox2}\,,\\
\rho_\mathrm{rad}(\hat{a})&={\rho}_I(1-\nu)\frac{\hat{a}^{4(1-\nu)}}{[1+\hat{a}^{4(1-\nu)}]^2}\label{eq:infaprox3}\,,
\end{align}
where we have introduced the total energy density at the early start of the inflationary period: ${\rho}_I=\frac{3}{\kappa^2}\Tilde{H}_I^2$.  It is remarkable to observe that the initial point $a=0$ is nonsingular in this framework since the Hubble function and the energy densities are well-defined functions taking finite values at that point: $H(0)=\Tilde{H}_I$, $\rho_\mathrm{vac}({0})={\rho}_I=\frac{3}{\kappa^2}\Tilde{H}_I^2$ and $\rho_\mathrm{rad}(0)=0$. In other words, RVM-inflation is characterized by a non-singular de Sitter phase.

It is also interesting to note that once we leave inflation behind and enter deep into the radiation-dominated epoch, i.e.  when $\hat{a}\gg 1$ (or  $a\gg a_*$) we obtain  the following asymptotic behavior: $\rho_\mathrm{rad}(\hat{a})\simeq {\rho}_I(1-\nu) \hat{a}^{-4(1-\nu)}\simeq \rho_I a_*^4 \,a^{-4}$ for small $|\nu|\ll1$. This behavior is essentially coincident with the standard evolution  law of the radiation energy density:  $\rho_\mathrm{rad}=\rho_\mathrm{rad}^0{a}^{-4}$. This allows to derive a useful estimate for $a_*$ which depends in part on measured parameters:

\begin{equation}\label{eq:astar}
a_{*}\sim
\left(\Omega_{\rm rad}^0\,\frac{\rco}{\rho_I}\right)^{\frac{1}{4}}\,.
\end{equation}
We shall use  this relation later on to estimate the numerical value of $a_{*}$ once we assess the order of magnitude of $\rho_I$.
Notice also the following important point, which can be read off immediately from equations \eqref{eq:infaprox2} and \eqref{eq:infaprox3}. During the radiation-dominated epoch, the vacuum energy is suppressed by the small factor $\nu$ (i.e. $\rho_\mathrm{vac}/\rho_\mathrm{rad}\sim \nu$) and hence the primordial BBN period can proceed standard\cite{{Asimakis:2021yct}}.

The scale of inflation $\Tilde{H}_I$ in our QFT context can be determined by comparing equations \eqref{eq:VEDinf2l} and \eqref{eq:rvm1}. We find
\begin{equation}\label{eq:HIvalue}
H_I=\sqrt{\frac{2\pi}{\xi-\frac{1}{6}} }\ m_\mathrm{Pl}\,.
\end{equation}
The above scale is not well defined in the conformal limit since there would be no running of the VED to account for inflation. In addition, we cannot accommodate a minimal coupling ($\xi=0$) in the presence of a single scalar field. Therefore, we need a non-minimal coupling satisfying $\xi>1/6$. This is to be expected, since only if this condition is met, the coefficient of the leading power $H^4$ in Eq.\,\eqref{eq:VEDinf2l}  would be positive to trigger inflation. However, with multiple scalar fields with different non-minimal couplings $\xi_i$, that condition could be relaxed and some of them (but not all) could be vanishing.

Notice that Eq.\,\eqref{eq:HIvalue} can also be written in terms of the particle mass as follows: $H_I={m}/{\sqrt{\epsilon}}$ using Eq.\eqref{eq:epsilonparameter}. This shows that $H\gg m$ during inflation since $\epsilon\ll1$.  For example, if we take into account the estimate $\nueff\sim  10^{-4}-10^{-2}$ \cite{SolaPeracaula:2021gxi} obtained from fitting the low energy data (i.e. from the current cosmological observations) we find from \eqref{eq:nueffAprox2} that $\epsilon \sim 10^{-6}-10^{-4}$, which implies that $H/m\sim 100-1000$ during the inflationary period. The estimate on $\epsilon$ determines in turn the order of magnitude of the non-minimal coupling parameter: $\xi-1/6=2\pi\epsilon \left(\mpl/m\right)^2$.  For particle masses near the GUT scale, $m\lesssim M_X\sim 10^{16}$GeV, we  have $\xi={\cal O}(100-1000)$, so $H_I<\mpl$.  However, as we have indicated, the final result will depend on the multiplicity of fields in the particular GUT considered. Finally, from Eq.\,\eqref{eq:nuI} and the above results we find $\nu_I=\epsilon\ln(m^2/H_I^2)=\epsilon\ln\epsilon$, and hence we can estimate numerically that $|\nu_I|={\cal O}\left(10^{-5}-10^{-3}\right)$.

\begin{figure}[t]
  \begin{center}
      \resizebox{1.10\textwidth}{!}{\includegraphics{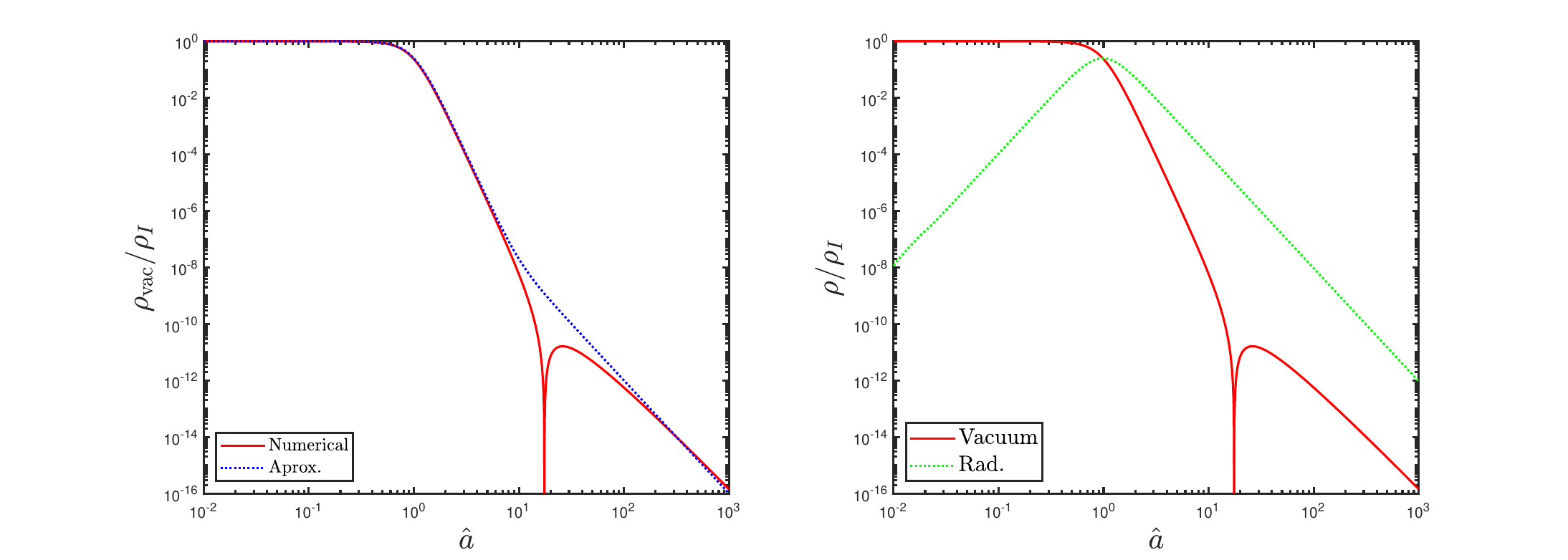}}
\caption{\textbf{(a)} Numerical solution of the VED  (solid line) versus the analytical solution \eqref{eq:infaprox2}(dotted line) around the transition time from inflation to the radiation epoch; \textbf{(b)} Numerical solution for both the VED and the radiation energy \eqref{eq:infaprox3} densities during the same transition period. In both cases $\nu=10^{-4}$, and $m$ of order of a GUT scale, the dependence being only logarithmic.}
\label{Fig:NumSolution}
  \end{center}
\end{figure}

Recall that the analytical solution \eqref{eq:infaprox} was based on solving the differential equation \eqref{eq:Hdiffeq} under the assumoption that $\nu=\nu_I=$const., whose numerical value we have just estimated.  This should be reasonable for the pure inflationary period ($\hat{a}<1$). However, when $\nu=\nu(H)$ evolves with the expansion, even as slowly  as in Eq.\eqref{eq:nuH}, we should expect some modification of the above picture. During inflation, $H\gg m$, but when inflation is over the Hubble rate can get close to $m$ and even substantially smaller. To check the impact of an evolving $\nu(H)$ in the immediate post-inflationary regime, e.g. when we enter the primordial radiation-dominated epoch, we need to solve Eq.\eqref{eq:Hdiffeq}  numerically.  We have used  Mathematica for that \cite{Mathematica}.  The result is shown in Fig. \ref{Fig:NumSolution}.
We only need to fix $m$ and $\epsilon$, which we have already performed in the context of a typical GUT,  and of course also a boundary condition for $H$ at the beginning of the radiation epoch when the inflationary period comes to an end. Since $|\nu|\ll 1$,  the analytical result (\ref{eq:infaprox}) can be used for the boundary condition of the numerical solution. Therefore, we impose that for $\hat{a}=1$:
\begin{equation}
    H(\hat{a}=1)\simeq \frac{1}{\sqrt{2}} H_I=\frac{1}{\sqrt{2}}\frac{m}{\sqrt{\epsilon}}\,.
\end{equation}
We note that for $\hat{a}\ll1$ the huge Hubble rate makes the system of equations insensible to $\nu$.  However, for $\hat{a}\sim100$ the VED has decayed enough into radiation and the precise value of $\nu(H)$ can play a role. Taking e.g. $\epsilon\sim 10^{-5}$ within our former estimate, the numerical solution yields $H(\hat{a}=100)\sim 10^{-5} m_\mathrm{Pl}$, and hence from \eqref{eq:nuH} we find $\nu \sim \mathcal{O}(10)\epsilon$ at this point. This contrasts to the situation where $\nu$ is fixed at the inflationary regime value \eqref{eq:nuI}, since in such a case $\nu_I$ even has the opposite sign to $\epsilon$ because $H\gg m$. When we move on to the low-energy stage, in which $H\ll m$, the sign of $\nu$ switches to that of $\epsilon$ and remains so until we approach the late universe when $\nu\to\nueff$ and Eq.\eqref{eq:nueffAprox2} holds.
\begin{figure}[t]
  \begin{center}
      \resizebox{0.53\textwidth}{!}{\includegraphics{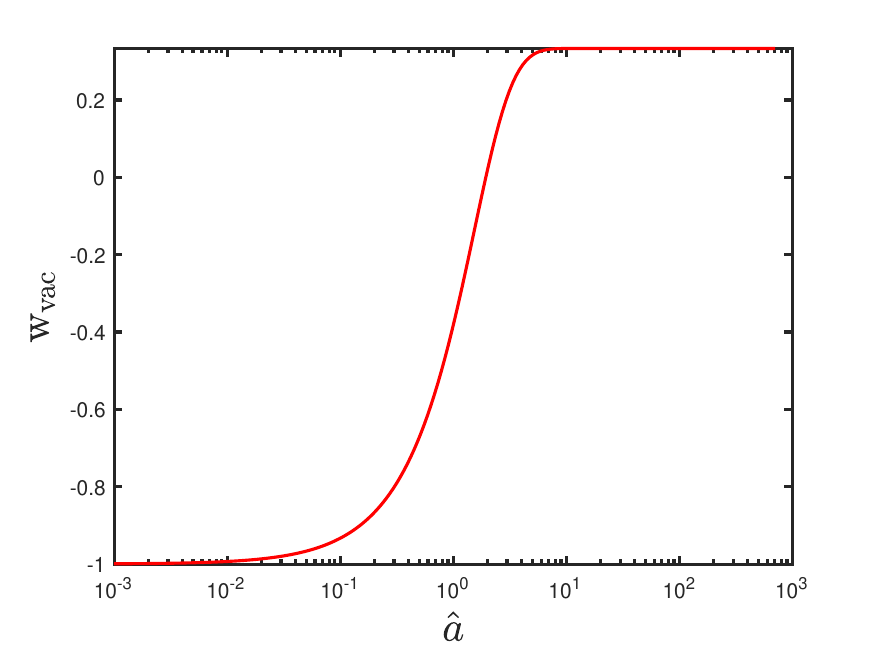}}
\caption{EoS of the running vacuum in the very early universe, as a function of $\hat{a}=a/a_*$. It is shown the transition from  the value at inflation  ($\wv=-1$) into de radiation-dominated epoch, where the vacuum adopts the EoS of radiation: $\wv=1/3$. Compare with Fig. \ref{Fig:EoSvacExtended}. }
\label{Fig:EoSvacuumInfl}
  \end{center}
\end{figure}

Let us now compare the numerical analysis of the energy densities in Fig. \ref{Fig:NumSolution} with the results of the analytical solution. From that figure we can see that in the beginning ($\hat{a}=0$) there is no radiation at all while the vacuum dominates and its energy density is the total energy density of the universe, $\rho_I$. This is also clearly reflected in the analytical solution for the vacuum and radiation energy densities, equations \eqref{eq:infaprox2} and \eqref{eq:infaprox3}. The vacuum then starts to decay very fast into radiation and  the inflationary period `graceful exits' into the standard FLRW radiation-dominated epoch. Moreover, after the universe exits the inflationary phase, which occurs for $\hat{a}\gg 1$, both energy densities scale approximately as $\rho\sim a^{-4}$, although the VED is
suppressed by a tiny coefficient $\nu$ -- which in this regime is actually the function $\nu(H)$, and hence this part can only be handled exactly through the numerical solution. The vacuum further evolves and its EoS changes from $-1$ during the inflationary period into $w_\mathrm{vac}\xrightarrow{}1/3$, i.e. adopting the same EoS as that of radiation, as previously noted from Eq.\eqref{eq:EoS2}. In the case of the very early universe, this can be clearly appreciated in Fig. \ref{Fig:EoSvacuumInfl}.

Overall, the inflationary epoch leads in a continuous way into the canonical RVM description of the late universe that we have studied in the previous section, see Eq.\,\eqref{eq:RVMcanonical}, which remains very close to the standard FLRW picture but possessing a mildly evolving VED as a fundamental distinctive feature. It is important to stress that the RVM account of the early cosmic history is capable not only of achieving a graceful exit from the inflationary era but also of leaving a remnant VED during the radiation-dominated epoch that is highly suppressed in front of the energy density of radiation. This fact is crucial in order not to spoil the  successful explanation  of the BBN  by the standard model of cosmology\,\cite{Asimakis:2021yct}. At the end of the day, it is reassuring to see that the analytical and numerical analyses of the cosmological transit from the inflationary phase into the standard radiation regime both lead to consistent conclusions.

\begin{figure}[t]
  \begin{center}
      \resizebox{0.58\textwidth}{!}{\includegraphics{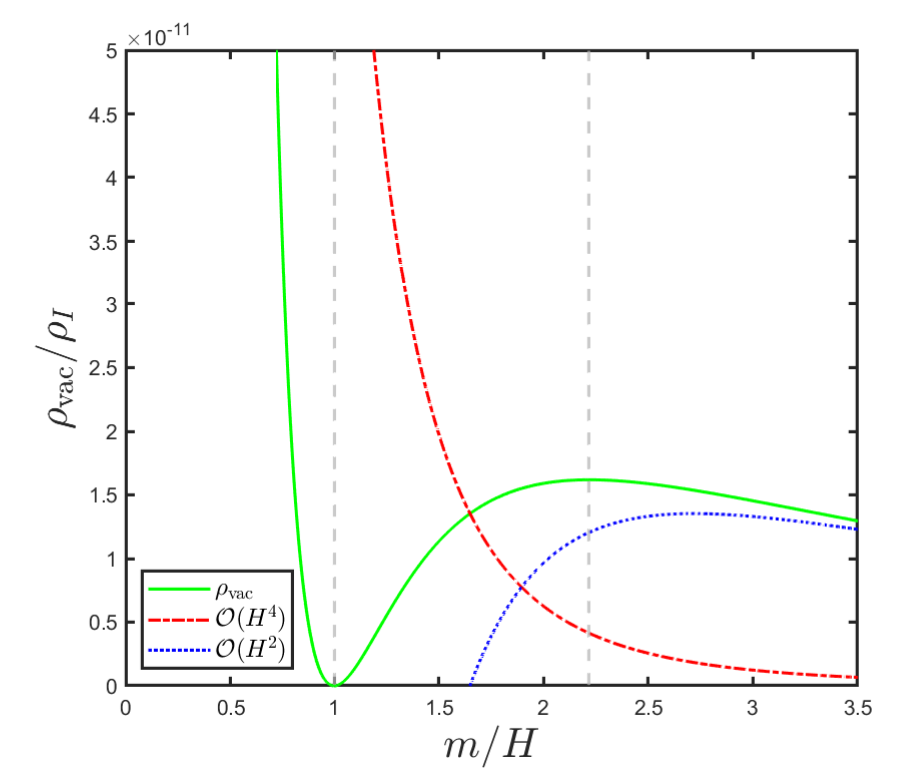}}
\caption{Numerical detail (in a normal scale) of the part of Fig. \ref{Fig:NumSolution} around the sharp dip produced by the logarithmic term of the VED in Eq.\,\eqref{eq:VEDinfl} once the early radiation epoch is attained and the $H^4$ power becomes subdominant versus the overall ${\cal O}(H^2)$ contribution. Shown are the full VED (solid line) and the individual contributions from the $H^4$ and $H^2$ terms (dotted lines).}
\label{Fig:NumDetails}
  \end{center}
\end{figure}

Some more details can be of interest, and we comment on them now briefly. For example, the corrections of order $\sim H^2$ become relevant only for the stage when the VED decreases. As can be seen in Fig. \ref{Fig:NumSolution}, the analytical solution proves to be quite reasonable in almost all relevant regimes, since it only misses the abrupt decay of the VED at the point $H=m$, which corresponds to a root of the equation $\rv(H)=0$, that is, we have $\rv(m)=0$, as is apparent from Eq.\,\eqref{eq:VEDinfl}. This root ultimate stems from our subtraction procedure (cf. Sec.\,\ref{ZPEScalar}) and is exact when we keep only the high energy terms of the VED, which are the only relevant ones in the early universe.  At this point, $\nu(H=m)=-\epsilon$, according to \eqref{eq:nuH}.  An amplification of this transition region for the VED from inflation into the radiation regime is displayed in Fig.\ref{Fig:NumDetails}. After the sharp trough in that figure, where the VED essentially vanishes, there is a local maximum that can be determined numerically and lies at $\hat{a}\simeq26.1$, a point that is already far from the inflationary stage. This is consistent with the fact that it corresponds to the point $m/H\simeq 2.218$ in the figure (marked with a vertical line) and hence at this point $H$ is already below the value of $m$. We also remark that despite the drastic transitory effect on the VED caused by the presence of the logarithm,  the impact on $\rho_\mathrm{rad}(a)$ and $H(a)$ is negligible since the analytical approximations previously found are virtually indistinguishable from the numerical solution, see Fig. \ref{Fig:NumSolution}(b). This is because at the time when $\rv(H)$ is sensitive to these transitory effects, the vacuum energy is already subdominant and the universe is well within the radiation-dominated epoch.

\subsection{Thermodynamical Aspects of $H^4$-Inflation }\label{sec:Thermoidynamics}

We should also note that the thermodynamic history of RVM-inflation is different from the conventional one associated with, say, an scalar inflaton. Strictly speaking there is no intermediate stage of (highly non-adiabatic) `reheating'\,\cite{KolbTurner}, characterized by superheavy massive particles decaying into conventional particles. In this sense, it is also different from the Starobinsky type of inflation\,\cite{Starobinsky:1980te}, in which, as already remarked, $H$ never goes through an inflationary regime characterized by $H=$const. and where, in addition,  there is an intermediate state of decay of the scalaron field.  Here, in contrast,  there is a continuous heating-up process caused by the decay of the quantum vacuum into relativistic particles.  The energy reservoir of RVM-inflation stems from the $H^4$-nonlinearities imprinted on the external gravitational field of the very early universe by the quantum fluctuations associated with the quantized matter fields. This is the ultimate driving force of RVM-inflation and the reason why the RVM has no horizon problem. Indeed, since $H\simeq H_I$ remains approximately constant during inflation, a light pulse beginning in
the remote past at $t=t_1\gtrsim t_i$ (i.e. shortly after inflation started at $t_i$) will have traveled until the end of inflation,
$t_f$, the physical distance\begin{equation}\label{eq:horizonInfinite}
d_{H}(a_f)= a(t_f)\int_{a_1}^{a_f}\frac{d
a'}{a'^2\,H}=\frac{a_f}{H_I}\left(\frac{1}{a_1}-\frac{1}{a_f}\right)\simeq
\left(\frac{a_f}{a_1}\right)\,H_I^{-1}\,.
\end{equation}
Since $a_f$ is  exponentially larger than
$a_1$ at the end of inflation, we have $a_f/a_1\ggg 1$ and so the
above integral (the particle horizon) can be as big as desired.
Therefore, all entropy production is causally produced, in contrast to the standard $\CC$CDM  model for which $d_H(a)/a\rightarrow 0$ for $a\to 0$ (and hence observers become isolated in the past).

The temperature of the heat bath generated from the decay of the primeval vacuum follows from equating the radiation density $\rho_{\rm rad}(a)$, given by Eq.\,\eqref{eq:infaprox3}, to the black-body formula $(\pi^2/30) g_*T^4$, where $g_*=\mathcal{O}(100)$ is the number of active degrees of freedom (d.o.f.) at the given temperature (e.g. $g_*=160.75$ in non-supersymmetric $SU(5)$). This yields
\begin{equation}\label{eq:TempRad}
T_\mathrm{rad}(\hat{a})=T_I\,(1-\nu)^{1/4} \frac{\hat{a}^{(1-\nu)}}{\left[1+\hat{a}^{4(1-\nu)}\right]^{1/2}}\,,
\end{equation}
where for convenience we have associated a temperature $T_I$ to the formerly defined total energy density through  $\rho_I=\frac{\pi^2}{30}\,g_\ast\,T_I^4$.
The maximum of the radiation temperature can be easily computed from Eq.\,\eqref{eq:TempRad}  and  is achieved at precisely the point $\hat{a}=1$ ($a=a_*$), with the following value:
\begin{equation}
T_\mathrm{max}=\frac{T_I}{\sqrt{2}}\,(1-\nu)^{1/4} \sim \left(\frac{45 m_\mathrm{Pl}^2 m^2}{16\pi^3g_* \epsilon} \right)^{1/4}=\left(\frac{45}{8\pi^2 g_{*}\left(\xi-1/6\right)}\right)^{1/4}\mpl\,,
\end{equation}
where in the above approximation we have neglected  terms of order $|\nu|\ll 1$. From the typical estimates on $g_*$ and $\xi$ mentioned above, it is easy to verify that this temperature remains one order of magnitude below the Planck mass, which is meaningful.  Notice that for $\hat{a}\gg1$ (i.e. $a\gg a_{*}$, corresponding to a region  deep into the radiation-dominated epoch) the scaling of the temperature \eqref{eq:TempRad} with the scale factor goes as
\begin{equation}
T_\mathrm{rad}\, a^{1-\nu}={\rm const.}
\end{equation}
Since $|\nu|\ll 1$,  we virtually recover the canonical scaling law of the adiabatic regime: $T_\mathrm{rad}\propto 1/a$.

Relevant for this thermodynamical discussion of RVM-inflation is certainly the issue of entropy production. Associated with the above radiation temperature, we may compute the corresponding (comoving) radiation entropy
$S_{r}= \left(4\rho_r/3T_{\rm rad}\right) a^3$\,\cite{KolbTurner}. The result is the following:
\begin{eqnarray}\label{eq:SrRVM}
S_{r}(\hat{a})=\frac{2\pi^2}{45}\,g_\ast T_{\rm rad}^3a^3=\frac{2\pi^2}{45}\,g_{*}\,T_I^3\,\astar^3\,(1-\nu)^{3/4}\,\frac{\ha^{6-3\nu}}{\Big[1+\ha^{4(1-\nu)}\Big]^{3/2}}\,.
\end{eqnarray}
From this formula we can see that during the heating-up period the comoving entropy rockets approximately as the sixth power of the scale factor, $S\sim \hat{a}^{(6-3\nu)}\sim \hat{a}^6$, until it finally reaches an approximate saturation plateau in the radiation-dominated phase:
\begin{equation}\label{eq:SrSaturation}
  S_r(\ha\gg1)\simeq \frac{2\pi^2}{45}\,g_{*}\,T_I^3\astar^3(1-\nu)^{3/4}\ha^{3\nu}
  \equiv S_{r0} \ha^{3\nu}\,.
  \end{equation}
It is not an exactly flat plateau for $\nu\neq 0$, but since $\nu$ is small the ulterior evolution of the entropy is much more tempered.
Equation \eqref{eq:SrSaturation} stands for  the (approximate)  asymptotic comoving entropy. For  $\nu=0$ the quantity $g_{*}T_{\rm rad}^3 a_r^3$ becomes
conserved during the adiabatic phase  and hence it must   equal the current value
$g_{s,0}\,T_{\gamma 0}^{3}\,a_0^{3}$, in which $T_{\gamma 0}\simeq 2.725\,$K
(CMB temperature now) and  $g_{s,0}=2+6\times
(7/8)\left(T_{\nu,0}/T_{\gamma 0}\right)^3\simeq 3.91$ is the entropy factor
for the light d.o.f. today, computed from the ratio of the present neutrino and photon temperatures. The upshot is that the huge entropy enclosed in our horizon today, $H_0^{-1}$, namely
\begin{equation}\label{eq:S0}
S_{0}=
\frac{2\pi^2}{45}\,g_{s,0}\,T_{\gamma 0}^3\,\left(H_0^{-1}\right)^3\simeq
2.3 h^{-3} 10^{87}\sim 10^{88} \ \ \ \ \ \ (h\simeq 0.7),
\end{equation}
can be explained from the approximately asymptotic value \eqref{eq:SrSaturation} acquired in
the radiation epoch after vacuum decay. This result cannot be accounted for in the standard model of cosmology without violating causality, this being the origin of the entropy and horizon problems\,\cite{KolbTurner}. In the present RVM framework, the large entropy generated at
the end of the inflation period is transferred to the radiation phase, and then it is preserved
by the standard (adiabatic) evolution, up to a small $\nu$-dependent correction. Thus, the observed entropy was causally produced in our remote past and the result does not depend
on the details of the underlying GUT.

If only within rough order of magnitude, it is easy to convince oneself from the above formulas that to match the desired total amount of entropy in the context of RVM-inflation, we need to  fulfill the condition  $T_I a_*\sim T_{\gamma 0}\sim 10^{-13}\,{\rm GeV}$. This relation  must be satisfied within order of magnitude so as to connect the inflationary epoch and the current epoch. It is a nontrivial condition in that  $T_{\gamma 0}$ is a measured quantity at present, whereas $T_I$ and $a_*$ are primordial parameters that belong to the very early universe and which we have previously estimated. Let us check that relation in an approximate way.  Upon using $\rho_I=3H_I^2\mpl^2/8\pi$ in combination with Eq.\,\eqref{eq:HIvalue}, and considering $\xi={\cal O}(10^{3})$ within the range of our estimate for this parameter, we find $\rho_I\sim 10^{73}$ GeV$^4$. Next using this result together with the measured quantities $\Omega_\mathrm{rad}^0\sim 10^{-4}$ and $\rho_c^0\sim 10^{-47}$ GeV$^4$ in   Eq.\,\eqref{eq:astar},  we can extract the following estimate for the inflationary scale $a_*$:
\begin{equation}\label{eq:astar2}
a_{*}\sim
\left(10^{-4}\,\frac{10^{-47}}{10^{73}}\right)^{1/4}\sim 10^{-31}\,.
\end{equation}
Putting the numbers together, we arrive at an estimate of the temperature at which the bulk of the radiation entropy was produced and subsequently leveled off: $T_I\sim T_{\gamma 0}/a_*\sim 10^{18}\,{\rm GeV}$, or equivalently  $T_I\sim 0.1 \mpl$, a result which is indeed in the ballpark of our original estimate for $T_I$. This shows the numerical consistency of our result within the rough order of magnitude. Notice that in performing the check we also utilized the range of  fitted values of $\nueff$, which was necessary to acquire an estimate for $\xi$. Of course, a more detailed calculation ought to take into account the multiplicity of fields in the theory and other considerations, but it is rewarding to see that the order of magnitude is nevertheless in place.

We should emphasize that the above discussion of RVM inflation has been derived within the general  QFT formulation of the running vacuum approach\cite{JSPRev2022}, which goes well beyond previous phenomenological considerations on these matters. Our approach provides a theoretical basis for a possible solution to these cosmological problems within fundamental physics. As noted in \cite{Sola:2015csa}, the clues to cosmological problems of the present may well have profound roots in the past.

Let us finally mention that although $H^4$-inflation is the simplest inflationary scenario in the context of the RVM, higher powers of the Hubble rate can also participate, for example $H^6/m^2$. These powers appear explicitly in the adiabatic expansion regardless of the value of the renormalization scale, although its calculation is more cumbersome; see\,\cite{CristianJoan2022a,CristianJoanSamira2023}.  However, $H^4$-inflation is the canonical mechanism for considering the transition to the radiation epoch. Furthermore, as noted in the Introduction, it has a stringy counterpart which is based on a process of Chern-Simons condensation in the very early universe\,\cite{ReviewNickJoan2021,PhantomVacuum2021}. In general, $H^4$-inflation in its various forms is very convenient as it allows to make contact with the exact de Sitter solution. This will be shown in a forthcoming publication.

\section{Discussion and Conclusions}\label{sec:conclusions}

In this work, we have further elaborated on the idea that the vacuum energy density (VED) in quantum field theory (QFT) in the expanding universe is a dynamical quantity which can provide a fundamental explanation not only for the dark energy (DE) but also for its time evolution throughout the cosmic history. This is the essence of the  approach called the running vacuum model (RVM), see \cite{JSPRev2022, JSPRev2013,JSPRev2015}  and references therein.  For more technical details on recent developments, see \cite{CristianJoan2020,CristianJoan2022a,CristianJoan2022b,CristianJoanSamira2023}.  Using the effective action approach, we have revisited the computation of the renormalized energy-momentum tensor (EMT) of a quantized scalar field non-minimally coupled to the FLRW background. We have derived the corresponding VED, and we have analyzed the consequences both for the current and for the very early universe.
The RVM approach indeed provides a nice unified picture of the cosmological evolution, which differs very little from the standard $\CC$CDM cosmological model in the late universe but provides a significant completion of the cosmic history at very early times; in particular, it describes the inflationary phase and its connection with the radiation-dominated epoch, a link which is completely missed by the $\CC$CDM model, wherein inflation is not described at all.

In our RVM context, we use an off-shell adiabatic renormalization prescription. As in any renormalization calculation in QFT, the renormalized VED in that context depends on a floating scale $M$. The existence of such a scale is characteristic of any renormalization scheme in QFT due to the intrinsic breaking of conformal invariance by quantum effects. The value of $M$ plays the role of renormalization point. The various epochs of the cosmic history (characterized by the value of the Hubble rate, $H$) are explored by setting that scale to the value of $H$ at each epoch.  The VED emerges as an expansion in an even number of time derivatives of the scale factor, which can be conveniently rephrased in terms of $H$ and its time derivatives, $\rho_{\rm vac}=\rho_{\rm vac}(H, \dot{H},\ddot{H},...)$.  The obtained expression for the VED can then be used to explore the entire history of the universe from the very early times where inflation occurs (triggered by the quantum effects associated to the quantized matter fields), going through the radiation- and matter-dominated epochs until reaching the current DE epoch. In this framework, what we call the dark energy density can be thought of as the remnant tail of the huge VED that brought about the exponential inflation at the very early times and is still decaying very slowly at present. Therefore, the RVM predicts in a natural way that the DE is dynamical.

In a number of phenomenological studies, the RVM has been successfully confronted with a large number of cosmological observations and it has been shown to seriously compete with the corresponding  $\CC$CDM description of the same data\,\cite{Sola:2015wwa,SolaPeracaula:2016qlq}. The currently preferred fitting values for the parameter $\nu_{\rm eff}$ that controls the running of the VED fall in the ballpark of $\sim 10^{-3}$ (depending on particular realizations, see  \cite{SolaPeracaula:2021gxi}). Such  analyses also show that the $H_0$-tension and the growth tension can both be significantly alleviated in the RVM context. These positive phenomenological implications notwithstanding,
perhaps the principal message of the RVM picture of the cosmic expansion is the following:  neither the VED at the present time, $\rho_{\rm vac}^0 =\rho_{\rm vac}(H_0)$, nor the associated cosmological `constant' $\Lambda=8\pi G\,\rho_{\rm vac}^0$ that we have measured, are really constants of nature in a fundamental QFT context.  For, at any expansion history time, the VED is given by the above mentioned function of $H$ and its time derivatives, and the associated  cosmological term is dynamical too: $\Lambda(H)=8\pi G(H)\,\rho_{\rm vac}(H)$. Needless to say,  the dynamics is smooth enough so as to make them appear as approximately constant and hence preserve the basic properties of the $\CC$CDM.   In other words,  the RVM behavior around the present time (and for that matter during the entire post-inflationary epoch)  is essentially $\Lambda$CDM-like. The VED change between two nearby epochs in recent history is $\delta \rho_{\rm vac}\propto \nu_{\rm eff}H^2$, where the small parameter $|\nu_{\rm eff}| \ll 1$ is calculable in QFT and plays the role of $\beta$-function of the VED running\,\cite{JSPRev2022}. Not less remarkable is the fact that the VED is free of the undesired quartic mass contributions $\sim m^4$  from any quantum matter field with a nonvanishing rest mass\,\cite{CristianJoan2020,CristianJoan2022a}. If they were present, these contributions would recreate the need for extreme fine-tuning, and hence would reproduce such an ugly feature of the  cosmological constant problem\,\cite{Weinberg89}.  Obviously, this theoretical property of the RVM is truly remarkable.

In the high-energy domain, i.e. for the very early universe,  the RVM provides a genuinely new mechanism of inflation based on a period where $H=$const.  The driving force of RVM-inflation is embodied in higher-order terms $H^{N}$ ($N\geq4$) that appear as additional quantum effects on the effective action of vacuum. These terms are irrelevant for the present universe, but play a major role in primeval cosmic times.  Here we have focused on the lowest-order even power that is capable of unleashing inflation in the very early universe, namely $H^4$. We found that it can describe the transition from the high-energy energy densities and entropy of the very early times into the standard FLRW regime. Therefore, it can implement the graceful exit of the inflationary phase into the radiation-dominated epoch and eventually leads to the current epoch.  In the process, the RVM can overcome the flatness and horizon problems as well as the entropy problem, which is a thermodynamic reformulation of the latter. We have checked that RVM-inflation can indeed account for the large entropy observed today in a way consistent with causality.

Last but not least, a potentially important implication of the RVM picture for the current universe (and hence subject to physical measurement) bears relation with the vacuum equation of state (EoS), which receives quantum corrections in our QFT context.  Despite that the background cosmology of the RVM is almost indistinguishable from the $\CC$CDM in the late universe, the vacuum EoS deviates slightly from $w=-1$ and can effectively mimic quintessence around the present time ($w_{\rm vac}(z)\gtrsim  -1$) or phantom DE ($w_{\rm vac}(z)\lesssim-1$), depending on the sign of $\nu_{\rm eff}$.  Interestingly, this result is actually consistent with generic parameterizations of the dynamical DE that have been revived from the recent release of DESI measurements\,\cite{DESI1,DESI2,DESI3}. Dynamical DE could be dynamical vacuum energy. It is to be noticed that our conclusions have been derived  within the framework of QFT in curved spacetime and therefore they might provide a fundamental explanation of dynamical DE in terms of running vacuum within a unified theory of the cosmological evolution.

\vspace{0.5cm}

{\bf Acknowledgments}:  JSP is funded by projects  PID2022-136224NB-C21 (MICIU), 2017-SGR-929 (Generalitat de Catalunya) and CEX2019-000918-M (ICCUB). AGF is funded by the project LCF/BQ/PI23/11970027. JSP and CMP acknowledge networking support by the COST Action CA21136
{\it Addressing observational tensions in cosmology
with systematics and fundamental physics (CosmoVerse)}. JSP also acknowledges networking support by  the COST Action CA23130 {\it Bridging high and low energies in search of quantum gravity (BridgeQG)}.

\end{document}